\documentclass[twocolumn,english,prl,preprintnumbers,amsmath,amssymb,superscriptaddress]{revtex4}
\usepackage[latin9]{inputenc}
\usepackage{amsmath}
\usepackage{graphicx}

\makeatletter
\@ifundefined{textcolor}{}
{%
 \definecolor{BLACK}{gray}{0}
 \definecolor{WHITE}{gray}{1}
 \definecolor{RED}{rgb}{1,0,0}
 \definecolor{GREEN}{rgb}{0,1,0}
 \definecolor{BLUE}{rgb}{0,0,1}
 \definecolor{CYAN}{cmyk}{1,0,0,0}
 \definecolor{MAGENTA}{cmyk}{0,1,0,0}
 \definecolor{YELLOW}{cmyk}{0,0,1,0}
}

\usepackage{dcolumn}
\usepackage{bm}
\usepackage{color}
\usepackage[final]{pdfpages}

\makeatother

\usepackage{babel}

\begin{document}

\preprint{preprint(\today)}

\title{Tunable nodal kagome superconductivity in charge ordered RbV$_{3}$Sb$_{5}$}

\author{Z.~Guguchia}
\email{zurab.guguchia@psi.ch} 
\affiliation{Laboratory for Muon Spin Spectroscopy, Paul Scherrer Institute, CH-5232 Villigen PSI, Switzerland}

\author{C.~Mielke III}
\affiliation{Laboratory for Muon Spin Spectroscopy, Paul Scherrer Institute, CH-5232 Villigen PSI, Switzerland}

\author{D.~Das}
\affiliation{Laboratory for Muon Spin Spectroscopy, Paul Scherrer Institute, CH-5232 Villigen PSI, Switzerland}

\author{R.~Gupta}
\affiliation{Laboratory for Muon Spin Spectroscopy, Paul Scherrer Institute, CH-5232 Villigen PSI, Switzerland}

\author{J.-X.~Yin}
\affiliation{Laboratory for Topological Quantum Matter and Advanced Spectroscopy (B7), Department of Physics,
Princeton University, Princeton, New Jersey 08544, USA}

\author{H.~Liu}
\affiliation{School of Physics and Astronomy, Key Laboratory of Artificial Structures and Quantum
Control (Ministry of Education), Shenyang National Laboratory for Materials Science, TsungDao Lee Institute, Shanghai Jiao Tong University, Shanghai 200240, China.}
\affiliation{Beijing National Laboratory for Condensed Matter Physics and Institute of Physics, Chinese
Academy of Sciences, Beijing 100190, China.}

\author{Q.~Yin}
\affiliation{Department of Physics and Beijing Key Laboratory of Opto-electronic Functional Materials and Micro-nano Devices, Renmin University of China, Beijing 100872, China}

\author{M.H.~Christensen}
\affiliation{Niels Bohr Institute, University of Copenhagen, 2100 Copenhagen, Denmark}

\author{Z.~Tu}
\affiliation{Department of Physics and Beijing Key Laboratory of Opto-electronic Functional Materials and Micro-nano Devices, Renmin University of China, Beijing 100872, China}

\author{C.~Gong}
\affiliation{Department of Physics and Beijing Key Laboratory of Opto-electronic Functional Materials and Micro-nano Devices, Renmin University of China, Beijing 100872, China}

\author{N.~Shumiya}
\affiliation{Laboratory for Topological Quantum Matter and Advanced Spectroscopy (B7), Department of Physics,
Princeton University, Princeton, New Jersey 08544, USA}

\author{Ts.~Gamsakhurdashvili}
\affiliation{Laboratory for Muon Spin Spectroscopy, Paul Scherrer Institute, CH-5232 Villigen PSI, Switzerland}

\author{M.~Elender}
\affiliation{Laboratory for Muon Spin Spectroscopy, Paul Scherrer Institute, CH-5232 Villigen PSI, Switzerland}

\author{Pengcheng~Dai}
\affiliation{Department of Physics and Astronomy, Rice Center for Quantum Materials, Rice University, Houston, TX, USA}

\author{A.~Amato}
\affiliation{Laboratory for Muon Spin Spectroscopy, Paul Scherrer Institute, CH-5232 Villigen PSI, Switzerland}

\author{Y.~Shi}
\affiliation{Beijing National Laboratory for Condensed Matter Physics and Institute of Physics, Chinese
Academy of Sciences, Beijing 100190, China.}
\affiliation{University of Chinese Academy of Sciences, Beijing 100049, China.}

\author{H.C.~Lei}
\affiliation{Department of Physics and Beijing Key Laboratory of Opto-electronic Functional Materials and Micro-nano Devices, Renmin University of China, Beijing 100872, China}

\author{R.M.~Fernandes}
\affiliation{School of Physics and Astronomy, University of Minnesota, Minneapolis, MN 55455, USA}

\author{M.Z. Hasan}
\affiliation{Laboratory for Topological Quantum Matter and Advanced Spectroscopy (B7), Department of Physics,
Princeton University, Princeton, New Jersey 08544, USA}
\affiliation{Princeton Institute for the Science and Technology of Materials, Princeton University, Princeton, New Jersey 08540, USA}
\affiliation{Materials Sciences Division, Lawrence Berkeley National Laboratory, Berkeley, California 94720, USA}
\affiliation{Quantum Science Center, Oak Ridge, Tennessee 37831, USA}
\author{H.~Luetkens}
\email{hubertus.luetkens@psi.ch} 
\affiliation{Laboratory for Muon Spin Spectroscopy, Paul Scherrer Institute, CH-5232 Villigen PSI, Switzerland}

\author{R.~Khasanov}
\email{rustem.khasanov@psi.ch} 
\affiliation{Laboratory for Muon Spin Spectroscopy, Paul Scherrer Institute, CH-5232 Villigen PSI, Switzerland}


\maketitle

\textbf{Unconventional superconductors often feature competing orders, small superfluid density, and nodal electronic pairing. While unusual superconductivity has been proposed in the kagome metals $A$V$_{3}$Sb$_{5}$, key spectroscopic evidence has remained elusive. Here we utilize pressure-tuned (up to 1.85 GPa) and ultra-low temperature (down to 18 mK) muon spin spectroscopy to uncover the unconventional nature of superconductivity in RbV$_{3}$Sb$_{5}$. At ambient pressure, we detect an enhancement of the width of the internal magnetic field distribution sensed by the muon ensemble, indicative of time-reversal symmetry breaking charge order. Remarkably, the superconducting state displays nodal energy gap and a reduced superfluid density, which can be attributed to the competition with the novel charge order. Upon applying pressure, the charge-order transitions are suppressed, the superfluid density increases, and the superconducting state progressively evolves from nodal to nodeless. Once charge order is eliminated, we find a superconducting pairing state that is not only fully gapped, but also spontaneously breaks time-reversal symmetry. Our results point to unprecedented tunable nodal kagome superconductivity competing with time-reversal symmetry-breaking charge order and offer unique insights into the nature of the pairing state.}


 
\begin{figure*}[t!]
\centering
\includegraphics[width=0.91\linewidth]{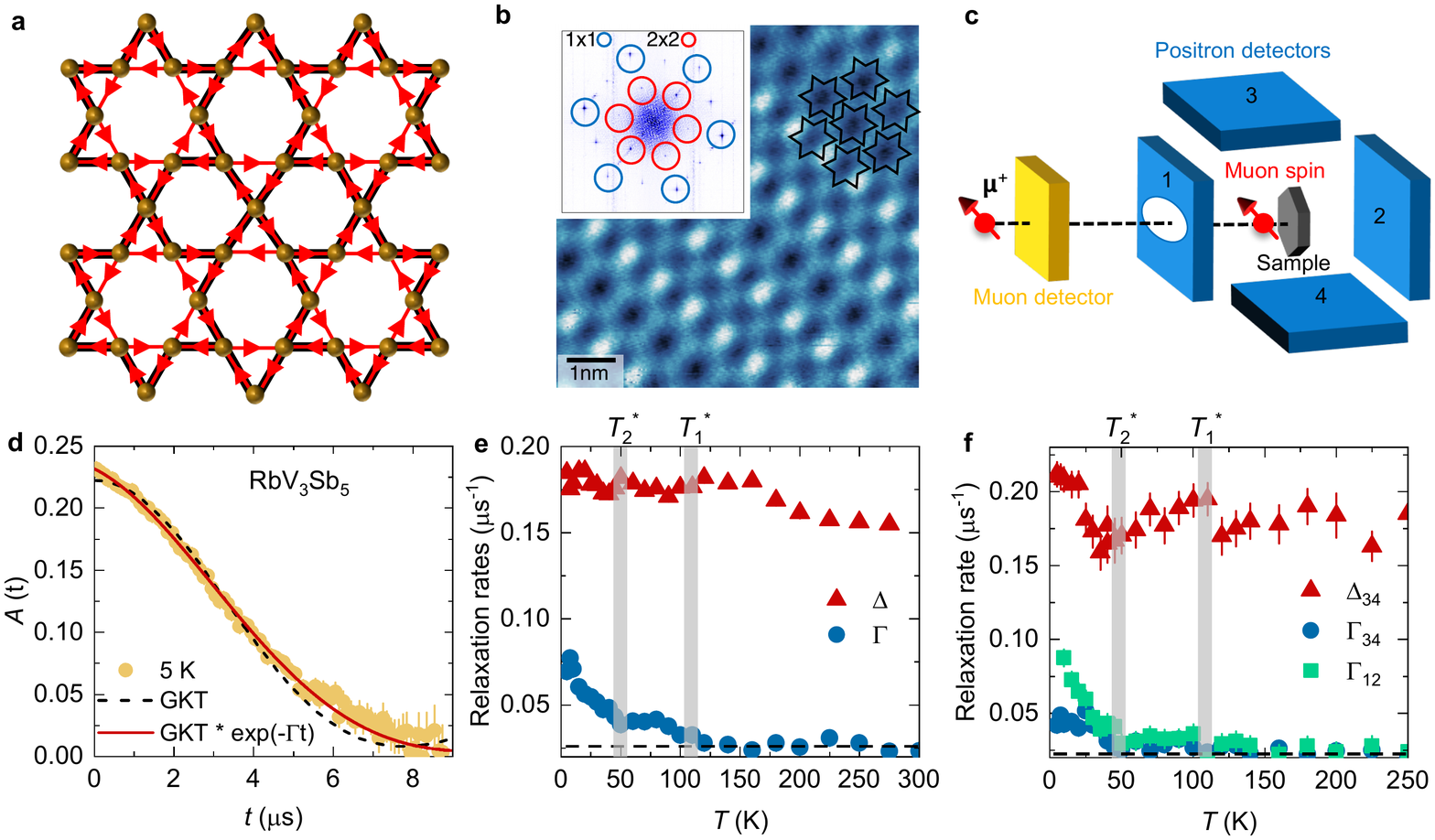}

\vspace{-2.5cm}

\includegraphics[width=0.91\linewidth]{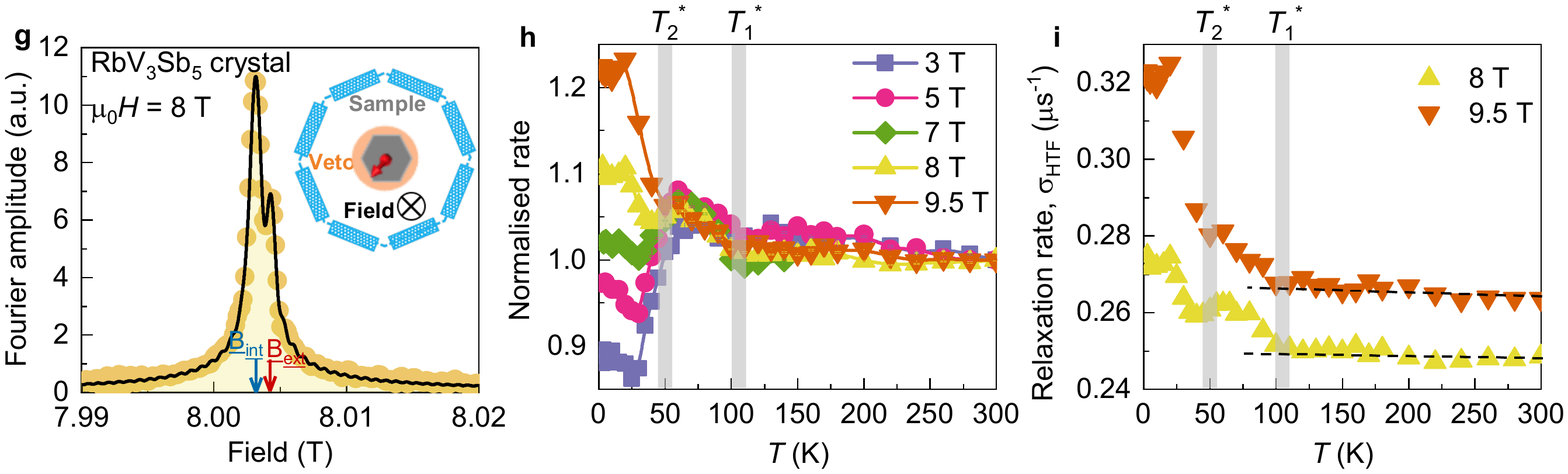}
\vspace{-6.0cm}
\caption{\label{fig:co_data} (Color online) \textbf{Time-reversal symmetry-breaking charge order in RbV$_{3}$Sb$_{5}$.} 
(a) Schematic example of an orbital current state (red arrows) in the kagome lattice. (b) Scanning tunneling microscopy of the Sb surface showing 2${\times}$2 charge order as illustrated by black lines. The inset is the Fourier transform of this image, displaying 1${\times}$1 lattice Bragg peaks (blue circles) and 2${\times}$2 charge-order peaks (red circles). The latter have different intensities, suggesting a chirality of the charge order. (c) A schematic overview of the experimental setup (see the methods section). (d) The ZF ${\mu}$SR time spectra for the polycrystalline sample of RbV$_{3}$Sb$_{5}$, obtained at $T$ = 5 K. The dashed and solid curves represent fits using the Gaussian Kubo Toyabe (GKT) function without (black) and with (red) a multiplied exponential $\rm{exp}(-\Gamma t)$ term, respectively. Error bars are the standard error of the mean (s.e.m.) in about 10$^{6}$ events. The temperature dependences of the relaxation rates ${\Delta}$ and ${\Gamma}$, which can be related to the nuclear and electronic contributions respectively, are shown in a wide temperature range for the polycrystalline (e) and the single crystal samples (f) of RbV$_{3}$Sb$_{5}$. Panel (f) presents ${\Gamma}$ obtained from two sets of detectors. The error bars represent the standard deviation of the fit parameters. (g) Fourier transform of the ${\mu}$SR asymmetry spectra for the single crystal of RbV$_{3}$Sb$_{5}$ at 3 K in the presence of an applied field of ${\mu}_{0} H = 8 $T. The black solid line is a two-component signal fit. The peaks marked by the arrows denote the external and internal fields, determined as the mean values of the field distribution from the silver sample holder (mostly) and from the sample, respectively. Inset shows the schematic high-field ${\mu}$SR experimental setup (see the methods section). (h) The temperature dependence of the high transverse field muon spin relaxation rate $\sigma_{\rm HTF}$ for the single crystal of RbV$_{3}$Sb$_{5}$, normalized to the value at 300 K, measured under different $c$-axis magnetic fields. 
(i) The temperature dependence of the relaxation rate, measured under magnetic field values of  ${\mu}$$_{0}H$ =  8 T and 9.5 T. }
\end{figure*}

 
  Due to their inherent geometric frustration and unique band structure, kagome materials \cite{Syozi} represent an excellent platform for discovering, classifying and understanding correlated electronic phases of quantum matter \cite{ZHou,GuguchiaCSS,JXYin3,CMielke,TbNature}.
The novel family of kagome metals $A$V$_{3}$Sb$_{5}$ ($A$ = K, Rb, Cs) \cite{BOrtiz2,JiangpingHu,TNeupert,QYin,YJiang} exhibit an array of interesting effects such as giant anomalous Hall conductivity \cite{SYang,FYu}, charge order
\cite{YJiang,MDenner,NShumiya,Wang2021,MKang,MHChristensen,MWenzel,YHu,Balents,Nandkishore}, orbital order \cite{DSong}, and possible unconventional superconductivity \cite{BOrtiz2,JiangpingHu,QYin}. An important feature of the charge order, which onsets at temperatures $T_{\rm co} \sim 100$ K at ambient pressure, is the breaking of time-reversal symmetry, as reported by scanning tunneling microscopy (STM) measurements \cite{YJiang,NShumiya,Wang2021,MKang} in (K,Rb,Cs)V$_{3}$Sb$_{5}$, by muon spin relaxation (${\mu}$SR) in KV$_{3}$Sb$_{5}$ \cite{GuguchiaMielke} and CsV$_{3}$Sb$_{5}$ \cite{LiYu}, and by Kerr effect measurements in CsV$_{3}$Sb$_{5}$ \cite{WuKerr}. 
This implies that the charge-ordered state displays not only bond distortions, but also orbital current loops (see Fig.~\ref{fig:co_data}{\bf a}) \cite{Haldane,Varma,NerJap,Fittipaldi}. 
  
Similarly to charge order, superconductivity, with transition temperature $T_c \sim 1$ K at ambient pressure, was also reported to display intriguing features, such as multiple gaps in (K,Cs)V$_{3}$Sb$_{5}$ \cite{HXu,NakayamaARPES,GuptaCVS}, diminished superfluid density in KV$_{3}$Sb$_{5}$ \cite{GuguchiaMielke}, and double-dome structures in the pressure phase diagrams of all three compounds \cite{KChen,NNWang,FDu}. However, no consensus on the superconducting gap structure has been reached yet \cite{GuguchiaMielke,YWang,HXu,NakayamaARPES,GuptaCVS,YGu,HChen,CZhao}, partly due to the challenges of performing spectroscopic studies under extreme conditions including ultra low temperatures and large pressures. Moreover, the role of the unconventional charge order in the emergence of these unusual superconducting features remains unclear, since the former onsets at a much higher temperature than the latter. In this regard, the sensitivity of both $T_c$ and $T_{\rm co}$ on applied pressure  \cite{KChen,NNWang,FDu} offers a rare setting to study the interplay between these two orders with a disorder-free tuning knob.   
 
Here, we tackle these issues by employing zero-field and high-field muon spin relaxation experiments to directly probe the interplay between charge order and superconductivity across the temperature-pressure phase diagram of RbV$_{3}$Sb$_{5}$. This allows us to assess not only the time-reversal symmetry-breaking nature of these two states, but also the evolution of the low-energy superconducting excitations as $T_{\rm co}$ is suppressed and $T_{\rm c}$ is enhanced. The latter measurements unearth a remarkable transition from nodal pairing, when superconductivity coexists with charge order, to nodeless pairing, when superconductivity onsets alone. They also reveal distinct relationships between $T_c$ and the superfluid density in these two regimes. The same behaviors are also observed in KV$_{3}$Sb$_{5}$, attesting to the robustness of our conclusions for the understanding of the pairing mechanism in the $A$V$_{3}$Sb$_{5}$ family. We discuss different scenarios for the symmetries of both the superconductivity and charge order states that may account for the unusual nodal-to-nodeless transition.


\section{Results and discussion}
\subsection{Probing spontaneous time-reversal symmetry breaking}

Scanning tunneling microscopy observes 2${\times}$2 charge order in RbV$_{3}$Sb$_{5}$ (Fig.~\ref{fig:co_data}$\bf{b}$ and Ref. \cite{NShumiya}) with an unusual magnetic field response \cite{NShumiya}, suggestive of time-reversal symmetry-breaking charge order. To directly probe signatures of time-reversal symmetry-breaking, we carried out zero field (ZF) ${\mu}$SR experiments (see Fig.~\ref{fig:co_data}{\bf c}) on both single crystal and polycrystalline samples of RbV$_{3}$Sb$_{5}$ above and below $T_{{\rm co}}$. The ZF-${\mu}$SR spectra (see Fig.~\ref{fig:co_data}{\bf d}) were fitted using the Gaussian Kubo-Toyabe (GKT) depolarization function \cite{Toyabe} multiplied with an exponential decay function \cite{GuguchiaMielke}:

\begin{equation}
\begin{aligned}
P_{ZF}^{\rm{GKT}}(t) =  \left(\frac{1}{3} + \frac{2}{3}(1 - \Delta^2t^2 ) \exp\Big[-\frac{\Delta^2t^2}{2}\Big]\right) \exp(-\Gamma t) \\ 
\end{aligned}
\end{equation}  
Here, ${\Delta}$/${\gamma_{\mu}}$ is the width of the local field distribution due to the nuclear moments and ${\gamma_{\mu}}$ = 0.085 $\mu$s$^{-1}$G$^{-1}$ is the muon gyromagnetic ratio. As we discussed previously in our work that reported time-reversal symmetry-breaking in KV$_{3}$Sb$_{5}$ \cite{GuguchiaMielke}, the exponential relaxation rate ${\Gamma}$ is mostly sensitive to the temperature dependence of the electronic contribution to the muon spin relaxation. In Fig.~\ref{fig:co_data}{\bf e}, the temperature dependence of the Gaussian and exponential relaxation rates ${\Delta}$ and ${\Gamma}$ for the polycrystalline sample of RbV$_{3}$Sb$_{5}$ are shown over a broad temperature range. The main observation is the two-step increase of the relaxation rate ${\Gamma}$, consisting of a noticeable enhancement at $T_{{\rm 1}}^{*}$ ${\simeq}$ 110 K, which corresponds approximately to the charge-order transition temperature $T_{{\rm co}}$, and a stronger increase below $T_{{\rm 2}}^{*}$ ${\simeq}$ 50 K. To substantiate this result, data from the single crystals are presented in Fig.~\ref{fig:co_data}{\bf f}. The data from the up-down (34) and forward-backward (12) sets of detectors not only confirm the increase of ${\Gamma}$, but also shed more light into the origin of the two-step behavior. In particular, while ${\Gamma}$$_{34}$ is enhanced mostly below $T_{{\rm 2}}^{*}$ ${\simeq}$ 50 K, ${\Gamma}$$_{12}$ also features a mild initial increase right below $T_{{\rm 1}}^{*}$ ${\simeq}$ 110 K. Since the enhanced electronic relaxation rate below $T_{{\rm 1}}^{*}$ is seen mostly in 
${\Gamma}$$_{12}$, it indicates that the local field at the muon site lies mostly within the ab-plane of the crystal. Below $T_{{\rm 2}}^{*}$, the internal field also acquires an out-of-plane component, as manifested by the enhancement of both ${\Gamma}$$_{12}$ and ${\Gamma}$$_{34}$. The increase of the electronic contribution to the internal field width is also accompanied by maxima and minima in the temperature dependence of the nuclear contribution to the internal field width ${\Delta}$/${\gamma_{\mu}}$, particularly for the up-down set of detectors (Figs.~\ref{fig:co_data}{\bf e} and {\bf f}).
 

The increase in the exponential relaxation of RbV$_{3}$Sb$_{5}$ between $T_{{\rm 1}}^{*}$ and 2 K is about 0.05 ${\mu}$$s^{-1}$, which can be interpreted as a characteristic field strength ${\Gamma}$$_{12}$/${\gamma_{\mu}}$ ${\simeq}$ 0.6~G. While these ZF-${\mu}$SR results are consistent with the onset of time-reversal symmetry-breaking at $T_{{\rm co}}$, high-field ${\mu}$SR experiments, illustrated in the inset of Fig.~\ref{fig:co_data}$\bf{g}$, are essential to confirm this effect, as we discussed previously \cite{GuguchiaMielke}. Fig.~\ref{fig:co_data}{\bf g} shows the probability distribution of the magnetic field measured at 3 K for the single crystal samples of RbV$_{3}$Sb$_{5}$ in the presence of a $c$-axis magnetic field of 8 T (see Methods for the details of the analysis). The contribution from the internal field is clearly identified. Fig.~\ref{fig:co_data}{\bf h} shows the corresponding temperature-dependent relaxation rate $\sigma_{\rm HTF}$ for different values of the external $c$-axis field. For 3 T, it displays a non-monotonic behavior, staying nearly constant across $T_{{\rm 1}}^{*}$ and then decreasing to a minimum before increasing again at low temperatures. Upon increasing the external field, the relaxation rate not only shows an increase right at $T_{{\rm 1}}^{*}$ ${\simeq}$ 110 K, but its temperature dependence below $T_{{\rm 2}}^{*}$ is reversed from being reduced to being enhanced upon lowering the temperature. Thus, as shown in Fig.~\ref{fig:co_data}, the relaxation rate extracted from the high-field ${\mu}$SR data shows a qualitatively similar two-step increase as the ZF data at the same characteristic temperatures $T_{{\rm 1}}^{*}$ ${\simeq}$ 110 K and $T_{{\rm 2}}^{*}$ ${\simeq}$ 50 K -- although the features are more pronounced at high fields. Because the temperature dependence of the nuclear contribution to the relaxation cannot be changed by an external field, we conclude that the two-step increase in the relaxation rate is driven by the electronic/magnetic contribution. 


\begin{figure*}[t!]
\centering
\includegraphics[width=0.95\linewidth]{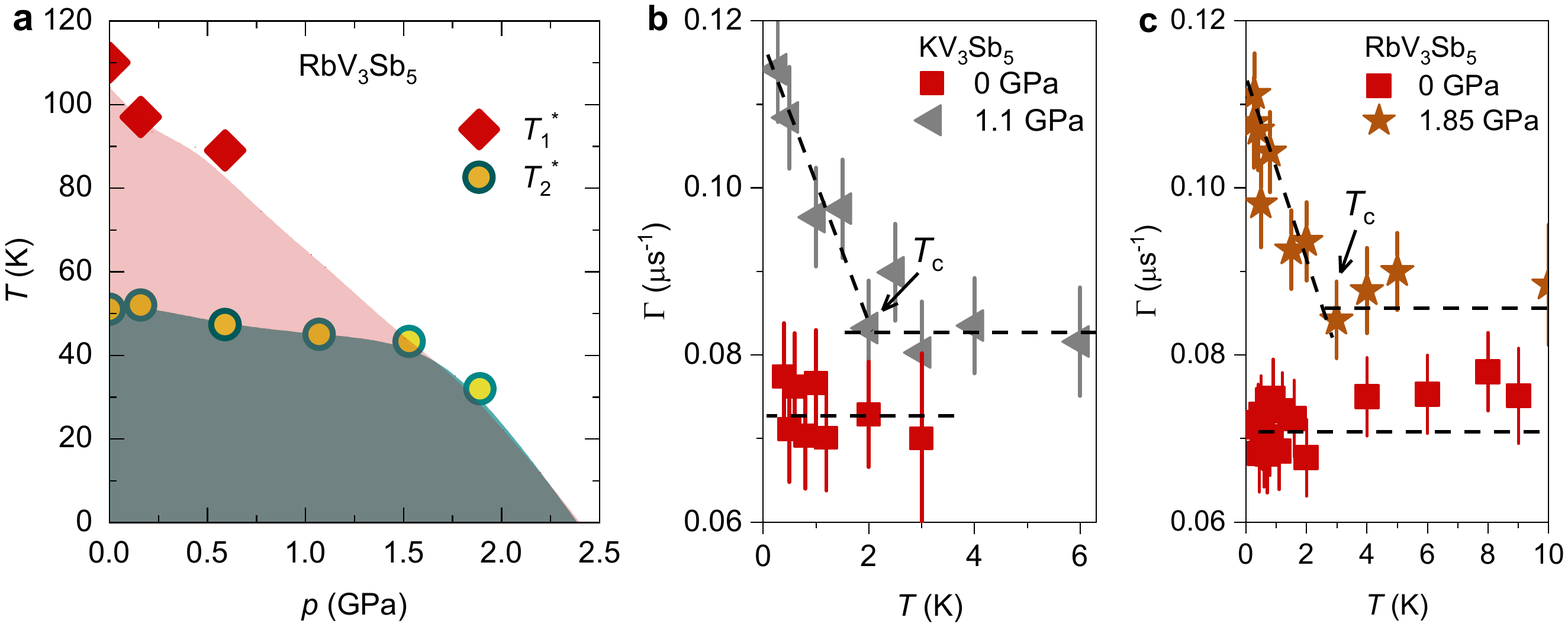} 
\vspace{-4.9cm}
\caption{\label{fig:sc_TRS} (Color online) \textbf{Time-reversal symmetry-breaking charge order and superconductivity in (K,Rb)V$_{3}$Sb$_{5}$ under pressure.} 
(a) The pressure dependences of the transition temperatures $T_{{\rm 1}}^{*}$ and $T_{{\rm 2}}^{*}$. Temperature dependence of the zero-field muon spin relaxation rate ${\Gamma}$ for KV$_{3}$Sb$_{5}$ (b) and RbV$_{3}$Sb$_{5}$ (c) in the temperature range across $T_{\rm c}$, measured at ambient pressure and above the critical pressure at which $T_{\rm c}$ is maximum. The error bars represent the standard deviation of the fit parameters.}
\end{figure*}

Therefore, the combination of ZF-${\mu}$SR and high-field ${\mu}$SR results on RbV$_{3}$Sb$_{5}$ provide direct evidence for time-reversal symmetry-breaking below the onset of charge order, which approximately coincides with $T_{{\rm 1}}^{*}$ ${\simeq}$ 110 K. As we previously discussed for KV$_{3}$Sb$_{5}$ \cite{GuguchiaMielke}, one plausible scenario to explain this effect is that loop currents along the kagome bonds are generated by a complex charge order parameter \cite{MDenner,Balents,Nandkishore}. Within this framework, muons can couple to the fields generated by these loop currents, resulting in an enhanced internal field width sensed by the muon ensemble (see also the Supplementary Information). The lower-temperature increase of the relaxation rate at $T_{{\rm 2}}^{*}$ ${\simeq}$ 50 K is suggestive of another ordered state that modifies such loop currents. An obvious candidate is a secondary charge-ordered state onsetting at $T_{{\rm 2}}^{*}$. Indeed, experimentally, it has been reported that some kagome metals may display two charge-order transitions \cite{MWenzel,YHu}. Theoretically, different charge-order instabilities have been found in close proximity \cite{MHChristensen}. Because 
time-reversal symmetry is already broken at $T_{{\rm 1}}^{*}$, it is not possible to distinguish, with our ${\mu}$SR data, whether this secondary charge-order state would break time-reversal symmetry on its own, or whether it is a more standard type of bond-charge-order. As shown in Figure \ref{fig:sc_TRS}{\bf a}, both $T_{{\rm 1}}^{*}$ and $T_{{\rm 2}}^{*}$ are suppressed by hydrostatic pressure. More specifically, the two-step charge-order transition becomes a single time-reversal symmetry-breaking charge-order transition at ${\sim}$ 1.5 GPa, above which $T_{{\rm 1}}^{*}$ = $T_{{\rm 2}}^{*}$ shows a faster suppression (see the Methods section for details).

\begin{figure*}[t!]
\centering
\includegraphics[width=1.0\linewidth]{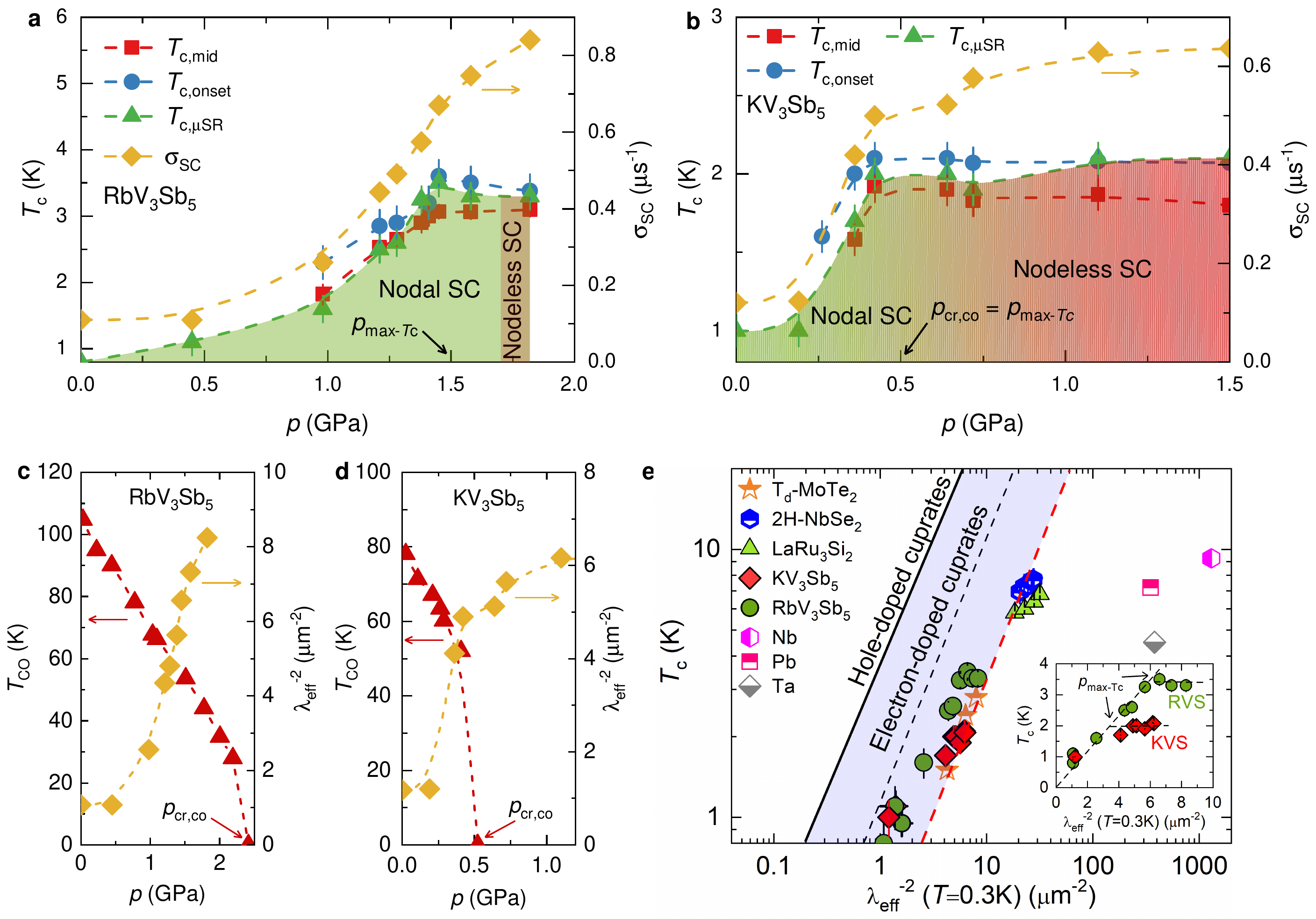}
\vspace{-0.5cm}
\caption{\label{fig:phase_diagrams} (Color online) \textbf{Coupled charge order and nodal superconductivity in kagome lattice.} 
Pressure dependence of the superconducting transition temperature (left axis) and of the base-$T$ value of ${\sigma}_{\rm sc}$ (right axis) for the polycrystalline samples of RbV$_{3}$Sb$_{5}$ (a) and KV$_{3}$Sb$_{5}$ (b). Here, $T_{\rm c,ons}$ and $T_{\rm c,mid}$ were obtained from AC measurements and $T_{\rm c,\mu SR}$, from $\mu$SR. Pressure dependence of  ${\lambda}_{\rm eff}^{-2}$ and charge order temperature $T_{\rm co}$ (Ref. \cite{NNWang}) for RbV$_{3}$Sb$_{5}$ (c) and KV$_{3}$Sb$_{5}$ (Ref. \cite{FDu}) (d). The arrows mark the critical pressure $p_{\rm cr,co}$ at which charge order is suppressed and the pressure $p_{\rm max-Tc}$ at which $T_{\rm c}$ reaches its maximum value. (e) Plot of $T_{\rm c}$ versus ${\lambda}_{\rm eff}^{-2}(0)$ in logarithmic scale obtained from our ${\mu}$SR experiments in KV$_{3}$Sb$_{5}$ and RbV$_{3}$Sb$_{5}$. Inset shows the plot in a linear scale. The dashed red line represents the relationship obtained for the kagome superconductor LaRu$_{3}$Si$_{2}$ as well as for the layered transition metal dichalcogenide superconductors $T_{d}$-MoTe$_{2}$ and 2H-NbSe$_{2}$ \cite{GuguchiaNbSe2,GuguchiaNature}. The relationship observed for cuprates is also shown \cite{Uemura1}, as are the points for various conventional superconductors . The error bars represent the standard deviation of the fit parameters.}
\end{figure*}

The same ZF-${\mu}$SR analysis can also be employed to probe whether there is time-reversal symmetry-breaking inside the superconducting state. Because charge order already breaks time-reversal symmetry at $T_{\rm co} \gg T_c$, it is necessary to suppress $T_{\rm co}$, which can be accomplished with pressure. The maximum pressure we could apply ($1.85$ GPa) is not enough to completely suppress the charge-order in RbV$_{3}$Sb$_{5}$, but it allows to enter the optimal $T_{\rm c}$ region of the phase diagram (see Fig. \ref{fig:phase_diagrams}{\bf a}) at which only a single time-reversal symmetry-breaking charge order transition is observed (see Fig. \ref{fig:sc_TRS}{\bf a}). This pressure value is also large enough to assess the pure superconducting state of the related compound KV$_{3}$Sb$_{5}$. In Fig. \ref{fig:sc_TRS}{\bf b}, we show the behavior of the internal field width $\Gamma$, extracted from the ZF-${\mu}$SR data, across the superconducting transition of KV$_{3}$Sb$_{5}$ measured both at ambient pressure (red, where charge-order is present) and at $1.1$ GPa (grey, where charge-order is absent). While at ambient pressure $\Gamma$ is little affected by superconductivity, at the higher pressure there is a significant enhancement of $\Gamma$, comparable to what has been observed in superconductors that are believed to spontaneously break time-reversal symmetry, such as SrRu$_{2}$O$_{4}$ \cite{LukeTRS}. The similar enhancement of $\Gamma$ below $T_{\rm c}$ $\sim$ 3 K is observed for RbV$_{3}$Sb$_{5}$ at $p$ = 1.85 GPa, as shown in Fig. \ref{fig:sc_TRS}{\bf c}. This provides strong evidence for time-reversal symmetry-breaking superconducting states in KV$_{3}$Sb$_{5}$ and RbV$_{3}$Sb$_{5}$, indicative of an unconventional pairing state.

\begin{figure*}[t!]
\centering
\includegraphics[width=1.0\linewidth]{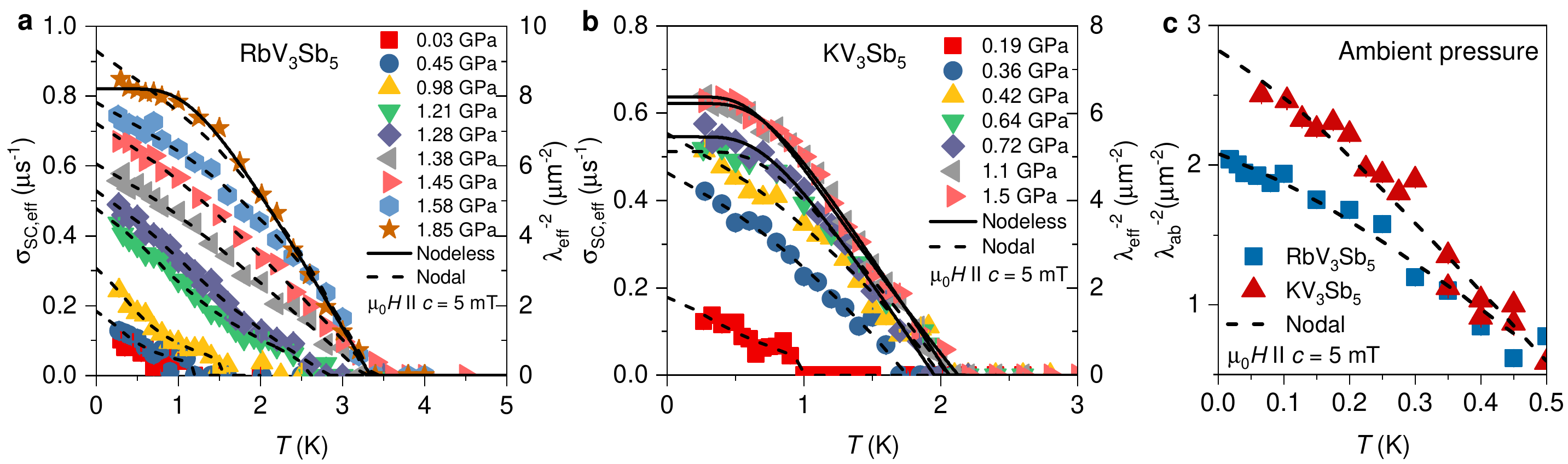}
\vspace{-0.5cm}
\caption{\label{fig:sc_data} (Color online) \textbf{Tunable nodal kagome superconductivity.} 
The temperature dependence of the superconducting muon spin depolarization rates ${\sigma}_{\rm sc}$ for RbV$_{3}$Sb$_{5}$ (a) and KV$_{3}$Sb$_{5}$ (b), measured in an applied magnetic field of ${\mu}_{\rm 0}H = 5$~mT at ambient and various applied hydrostatic pressures. The error bars represent the standard deviations of the fit parameters. The solid (dashed) lines correspond to a fit using a model with nodeless (nodal) two-gap superconductivity. (c) The inverse squared penetration depth ${\lambda}_{ab}^{-2}$ for the single crystals of KV$_{3}$Sb$_{5}$ and RbV$_{3}$Sb$_{5}$ as a function of temperature at ambient pressure.}
\end{figure*}

\subsection{Superfluid density as a function of pressure}

An additional property of the superconducting state that can be directly measured with $\mu$SR is the superfluid density. This is accomplished by extracting the second moment of the field distribution from the muon spin depolarization rate ${\sigma}_{{\rm sc}}$, which is related to the superconducting magnetic penetration depth $\lambda$ as $\left<\Delta B^2\right>\propto \sigma^2_{sc} \propto \lambda^{-4}$ (see Methods section). Because $\lambda^{-2}$ is proportional to the superfluid density, so is $\sigma_{sc}$. Figures~\ref{fig:phase_diagrams} and \ref{fig:sc_data} summarise the pressure and temperature dependences of ${\sigma}_{{\rm sc}}$ (measured in an applied magnetic field of ${\mu}_{\rm 0}H = 5$~mT) in the superconducting states of RbV$_{3}$Sb$_{5}$ and KV$_{3}$Sb$_{5}$. As the temperature is decreased below $T_{{\rm c}}$, the depolarization rate ${\sigma}_{{\rm sc}}$ starts to increase from zero due to the formation of the flux-line lattice (see Fig.~\ref{fig:sc_data}{\bf a}). As the pressure is increased, not only $T_{{\rm c}}$ (as determined from AC susceptibility and ${\mu}$SR experiments), but also the low-temperature value of ${\sigma}_{{\rm sc}}$ (measured at the baseline of 0.25 K) show a substantial increase for both compounds, as shown in Figs.~\ref{fig:phase_diagrams}{\bf a} and {\bf b}. In both cases, $T_{\rm c,ons}$ first quickly reaches a maximum at a characteristic pressure $p_{\rm max-Tc}$, namely, 3.5 K at $p_{\rm max-Tc}$ ${\simeq}$ 1.5 GPa for the Rb compound and 2.3 K at $p_{\rm max-Tc}$ ${\simeq}$ 0.5 GPa for the K compound. Beyond those pressure values, the transition temperature remains nearly unchanged. The superfluid density ${\sigma}_{{\rm sc}}$(0.25 K) also increases significantly from its ambient-pressure value upon approaching $p_{\rm max-Tc}$, by a factor of approximately $7$ for the Rb compound and $5$ for the K system. In both cases, ${\sigma}_{{\rm sc}}$(0.25 K) continues increasing beyond $p_{\rm max-Tc}$, although at a lower rate that may indicate approach to saturation.  


These behaviors are consistent with competition between charge order and superconductivity. Indeed, as shown in Figs.~\ref{fig:phase_diagrams}{\bf c} and {\bf d}, the increase in the superfluid density is correlated with the decrease in the charge ordering temperature $T_{\rm co}$. More specifically, the pressure values $p_{\rm max-Tc}$ for which $T_c$ is maximum are close to the critical pressures $p_{\rm cr,co}$ beyond which charge order is completely suppressed. In fact, as displayed in Fig.~\ref{fig:phase_diagrams}{\bf b},  $p_{\rm cr,co}$ essentially coincides with $p_{\rm max-Tc}$ for KV$_{3}$Sb$_{5}$. Competition with charge order could naturally account for the suppression of the superfluid density towards the low-pressure region of the phase diagram, where $T_{\rm co}$ is the largest. Since charge order partially gaps the Fermi surface, as recently seen by quantum oscillation \cite{OrtizPRX} and ARPES \cite{NakayamaARPES,MKang} studies, the electronic states available for the superconducting state are suppressed, thus decreasing the superfluid density \cite{Machida1984,Fernandes2010}.

Having extracted ${\sigma}_{{\rm sc}}$, we can directly obtain the magnetic penetration depth $\lambda$ (see Methods). For polycrystalline samples, this gives an effective penetration depth $\lambda_{\rm eff}$, whereas for single crystals, it gives the in-plane $\lambda_{ab}$. It is instructive to plot the low-temperature penetration depth not as a function of pressure, but as a function of  $T_c$  \cite{Uemura1}. As shown in Fig.~\ref{fig:phase_diagrams}{\bf e}, the ratio $T_{\rm c}/\lambda_{\rm eff}^{-2}$ for unpressurized RbV$_{3}$Sb$_{5}$ is ${\sim}$ 0.7, similar to the one previously reported for KV$_{3}$Sb$_{5}$ \cite{GuguchiaMielke}. This ratio value is significantly larger from that of conventional BCS superconductors, indicative of a much smaller superfluid density. Moreover, we also find an unusual relationship between $\lambda_{\rm eff}^{-2}$ and $T_{\rm c}$ in these two kagome superconductors, which is not expected for conventional superconductivity. This is presented in the inset of Fig.~\ref{fig:phase_diagrams}{\bf e}: below $p_{\rm max-Tc}$, the superfluid density (which is proportional to $\lambda_{\rm eff}^{-2}$) depends linearly on $T_{\rm c}$, whereas above $p_{\rm max-Tc}$, $T_{c}$ barely changes for increasing $\lambda_{\rm eff}^{-2}$. Historically, a linear increase of $T_{\rm c}$ with $\lambda_{\rm eff}^{-2}$ has been observed only in the underdoped region of the phase diagram of unconventional superconductors. Deviations from linear behavior were previously found in an optimally doped cuprate \cite{GuguchiaNbSe2}, in some Fe-based superconductors \cite{GuguchiaNature}, and in the charge-ordered superconductor 2H-NbSe$_{2}$ under pressure \cite{GuguchiaNbSe2}. Therefore, in RbV$_{3}$Sb$_{5}$ and KV$_{3}$Sb$_{5}$, it is tempting to attribute this deviation to the suppression of the competing charge ordered state by the applied pressure. More broadly, these two different dependences of $T_{\rm c}$ with $\lambda_{\rm eff}^{-2}$ indicate superconducting states with different properties below and above $p_{\rm max-Tc}$.

To further probe this scenario, we quantitatively analyze the temperature dependence of the penetration depth ${\lambda}(T)$ \cite{GuguchiaMoTe2} for both compounds as a function of pressure, see Figs.~\ref{fig:sc_data}{\bf a} and {\bf b}. Quite generally, upon decreasing the temperature towards zero, a power-law dependence of $\lambda_{\rm eff}^{-2}(T)$ is indicative of the presence of nodal quasiparticles, whereas an exponential saturation-like behavior is a signature of a fully gapped spectrum. The low-$T$ behavior of $\lambda_{ab}^{-2}(T)$ for single crystals of RbV$_{3}$Sb$_{5}$ and KV$_{3}$Sb$_{5}$, measured down to 18 mK and shown in Fig.~\ref{fig:sc_data}{\bf c}, displays a linear-in-$T$ behavior, consistent with the presence of gap nodes. Quantitatively, the curve is well described by a phenomenological two-gap model, where one of the gaps has nodes and the other does not (see Methods). 

Such a linear-in-$T$ increase of $\lambda_{\rm eff}^{-2}(T)$ upon approaching $T=0$ is also seen in polycrystalline samples for pressures up to $p_{\rm max-Tc}$. In the case of RbV$_{3}$Sb$_{5}$ (Fig.~\ref{fig:sc_data}{\bf a}), for the only pressure value available above $p_{\rm max-Tc} \approx 1.5$ GPa, the penetration depth curve seems to be better fitted by a model with a nodeless gap. This is much clearer in the case of KV$_{3}$Sb$_{5}$ (Fig.~\ref{fig:sc_data}{\bf b}): above $p_{\rm max-Tc} \approx 0.5$ GPa, $\lambda_{\rm eff}^{-2}(T)$ displays a saturation-like behavior that is well captured quantitatively by a model with a nodeless gap. Since $p_{\rm max-Tc}$ is close to $p_{\rm co,cr}$, specially for the K compound, these results show that charge order strongly influences the superconducting gap structure in (Rb,K)V$_{3}$Sb$_{5}$, inducing nodes in an otherwise fully gapped pairing state. To the best of our knowledge this is the first direct experimental demonstration of a plausible pressure-induced change in the superconducting gap structure from nodal to nodeless in these kagome superconductors. 

One possible explanation for these results is on the changes that the emergence of charge order causes on the Fermi surface. First-principle calculations on AV$_{3}$Sb$_{5}$ compounds indicate the existence of multiple Fermi pockets in the absence of charge order \cite{OrtizPRX}. The simplest fully-gapped pairing state is an s-wave one consisting of different nodeless gaps (with potentially different signs) around each pocket. The onset of long-range charge order further breaks up these pockets into additional smaller pockets. Depending on the relative sign between the original gaps and on the details of the reconstructed Fermi pockets, accidental nodes could emerge. Such a scenario was proposed in the case of competing $s^{+-}$-wave superconductivity and spin-density wave in iron-pnictide superconductors \cite{Maiti2012}.

The main drawback of this scenario is that it does not account for the time-reversal symmetry-breaking of the ``pure" superconducting state. In this regard, a fully gapped pairing state that also breaks time-reversal symmetry is the chiral $d_{x^2-y^2} + id_{xy}$ state \cite{Kiesel,XWu}. As long as the charge ordered state preserves the point-group symmetry of the disordered state, the chiral pairing symmetry is expected to be retained below $T_{\rm co}$, suggesting a nodeless superconducting state. However, if the charge-ordered state breaks the threefold rotational symmetry of the lattice, as proposed experimentally \cite{Xiangnematic} and theoretically \cite{Balents,MHChristensen} for certain AV$_{3}$Sb$_{5}$ compounds, a nodal gap is stabilized for a sufficiently large charge order parameter, as we show in the supplementary material. In this case, the nodal-to-nodeless transition does not coincide with the full suppression of charge order, unless the transition from the charge-ordered superconducting state to the ``pure" superconducting state is first-order. We note that the same conclusions would also apply for the triplet chiral $p_{x} + ip_{y}$ state.

\section{Conclusion}

 Our results provide direct evidence for unconventional superconductivity in (Rb,K)V$_{3}$Sb$_{5}$, by combining the observations of nodal superconducting pairing and a small superfluid density at ambient pressure, which in turn displays an unconventional dependence on the superconducting critical temperature. Moreover, we find that the hydrostatic pressure induces a change from a nodal superconducting gap structure at low pressure to a nodeless, time-reversal symmetry-breaking superconducting gap structure at high pressure, a behavior correlated with the suppression of time-reversal symmetry-breaking charge order. Our results point to the rich interplay and accessible tunability between nodal unconventional superconductivity and time-reversal symmetry-breaking charge orders in the correlated kagome lattice, offering new insights into the microscopic mechanisms involved in both orders.


\newpage 

\section{METHODS}

\textbf{Experimental details}: Zero field (ZF) and transverse field (TF) $\mu$SR experiments at ambient pressure on the single crystalline and polycrystalline samples of RbV$_{3}$Sb$_{5}$ and KV$_{3}$Sb$_{5}$ were performed on the GPS instrument and high-field HAL-9500 instrument, equipped with BlueFors vacuum-loaded cryogen-free dilution refrigerator (DR), at the Swiss Muon Source (S$\mu$S) at the Paul Scherrer Institut, in Villigen, Switzerland. ${\mu}$SR experiments under pressure were performed at the ${\mu}$E1 beamline of the Paul Scherrer Institute (Villigen, Switzerland using the instrument GPD, where an intense high-energy ($p_{\mu}$ = 100 MeV/c) beam of muons is implanted in the sample through the pressure cell. The $^{4}$He cryostats equipped with the $^{3}$He insets (base temperature $\simeq0.25$~K) were used. A mosaic of several crystals stacked on top of each other was used for these measurements. The magnetic field was applied both in-plane (along the $ab$-plane) and out-of-plane (along the crystallographic $c$-axis). A schematic overview of the experimental setup for zero-field and low transverse field measurements is shown in Figure 1c. The muon spin is forming $45^{\circ}$ with respect to the $c$-axis of the crystal. The sample was surrounded by four detectors: Forward (1), Backward (2), Up (3) and Down (4). A schematic overview of the experimental setup for high-field $\mu$SR instrument is shown in the inset of Figure 1g. The muon spin forms $90^{\circ}$ with respect to the $c$-axis of the crystal. The sample was surrounded by 2 times 8 positron detectors, arranged in rings. The specimen was mounted in a He gas-flow cryostat with the largest face perpendicular to the muon beam direction, along which the external field was applied. Zero field and high transverse field $\mu$SR data analysis on single crystals of RbV$_{3}$Sb$_{5}$ were performed using both the so-called asymmetry and single-histogram modes \cite{Yaouanc_book_2011}.\\

\textbf{Sample growth}: Single crystals of RbV$_{3}$Sb$_{5}$ were synthesized by Rb ingot (purity 99.9 ${\%}$), V powder (purity 99.9 ${\%}$) and Sb grains (purity 99.999 ${\%}$) using the self-flux method \cite{QYin}.\\

\textbf{${\mu}$SR experiment}: In a ${\mu}$SR experiment \cite{Sonier} nearly 100 ${\%}$ spin-polarized muons ${\mu}$$^{+}$
are implanted into the sample one at a time. The positively
charged ${\mu}$$^{+}$ thermalize at interstitial lattice sites, where they
act as magnetic microprobes. In a magnetic material the 
muon spin precesses in the local field $B_{\rm \mu}$ at the
muon site with the Larmor frequency ${\nu}_{\rm \mu}$ = $\gamma_{\rm \mu}$/(2${\pi})$$B_{\rm \mu}$ (muon
gyromagnetic ratio $\gamma_{\rm \mu}$/(2${\pi}$) = 135.5 MHz T$^{-1}$). 
Using the $\mu$SR technique, important length scales of superconductors can be measured, namely the magnetic penetration depth $\lambda$ and the coherence length $\xi$. If a type II superconductor is cooled below $T_{\rm c}$ in an applied magnetic field ranging between the lower ($H_{c1}$) and the upper ($H_{c2}$) critical fields, a vortex lattice is formed which in general is incommensurate with the crystal lattice, with vortex cores separated by much larger distances than those of the crystallographic unit cell. Because the implanted muons stop at given crystallographic sites, they will randomly probe the field distribution of the vortex lattice. Such measurements need to be performed in a field applied perpendicular to the initial muon spin polarization (so-called TF configuration).\\

\textbf{Pressure cell}:  Pressures up to 1.9 GPa were generated in a double wall piston-cylinder type cell made of CuBe/MP35N, specially designed to perform ${\mu}$SR experiments under pressure \cite{KhasanovHPR}. A fully assembled typical double-wall pressure cell is presented
in Extended Data Fig. 1. The body of the pressure cell consists of two parts: the inner and the outer cylinders which are shrink fitted into each other. Outer body of the cell is made out of MP35N alloy. Inner body of the cell is made out of CuBe alloy. Other components of the cell are: pistons, mushroom, seals, locking nuts, and spacers.  The mushroom pieces and sealing rings were made out of non hardened Copper Beryllium. With both pistons completely inserted, the maximum sample height is 12 mm. As a pressure transmitting medium Daphne oil was used. The pressure was measured by tracking the superconducting transition of a very small indium plate by AC susceptibility. The filling factor of the pressure cell was maximized. The fraction of the muons stopping in the sample was approximately 40 ${\%}$.\\

\textbf{Crystal structure of RbV$_{3}$Sb$_{5}$}: Additional characterization information is provided here on the kagome superconductor RbV$_{3}$Sb$_{5}$ which crystallizes in the novel $A$V$_{3}$Sb$_{5}$-type structure (space group $P6$/$mmm$, where $A$ = K,~Rb,~Cs). The crystallographic structure of prototype compound RbV$_{3}$Sb$_{5}$ shown in panel (a) of Extended Data Figure 2 illustrates how the V atoms form a kagome lattice (medium beige circles) intertwined with a hexagonal lattice of Sb atoms (small red circles). The Rb atoms (large purple circles) occupy the interstitial sites between the two parallel kagome planes. In panel (b) the vanadium kagome net has been emphasized, with the interpenetrating antimony lattice included to highlight the unit cell (see dashed lines). Extended data Figures 2c shows an optical microscope image of several single crystals of RbV$_{3}$Sb$_{5}$ on millimeter paper. The Laue X-ray diffraction image (see the Extended data Figure 2d) demonstrates the single crystallinity of the samples used for ${\mu}$SR experiments.\\

\textbf{Magnetization measurements of RbV$_{3}$Sb$_{5}$}: The magnetization measurements show the abrupt drop in macroscopic magnetization 
across $T_{{\rm 1}}^{*}$ = $T_{\rm CDW,1}$ ${\simeq}$ 105 K for the field applied along the $c$-axis, as shown in Extended Figure 3.
Interestingly, a shallow minimum around $T_{{\rm 2}}^{*}$ = $T_{\rm CDW,2}$ ${\simeq}$ 50 K is also seen in magnetization, followed by sizeable increase at lower temperatures.\\

\textbf{Analysis of high field TF-${\mu}$SR data}: Figure 1g shows the probability field distribution, measured at 3 K for the single crystal samples of RbV$_{3}$Sb$_{5}$ in the $c$-axis magnetic field of 8 T. In the whole investigated temperature range, two-component signals were observed: a signal with fast relaxation (low frequency) and another one with a slow relaxation (high frequency). The narrow signal arises mostly from the muons stopping in the silver sample holder and its position is a precise measure of the value of the applied magnetic field. The width and the position of the narrow signal is assumed to be temperature independent and thus they were kept constant in the analysis. The relative fraction of the muons stopping in the sample was fixed to the value obtained at the base-$T$ and kept temperature independent. The signal with the fast relaxation, which is shifted towards the lower field from the applied one, arises from the muons stopping in the sample and it takes a major fraction (80 ${\%}$) of the ${\mu}$SR signal. This points to the fact that the sample response arises from the bulk of the sample. We note that in high magnetic fields we cannot systematically discriminate between the nuclear and the electronic contribution to the relaxation rate and thus we show the total high-field muon spin relaxation rate $\sigma_{\rm HTF}$ in Figure 1i.\\ 

\textbf{Knight shift of RbV$_{3}$Sb$_{5}$}: Extended Data Figure 4 shows the the temperature dependence of the Knight shift, measured at various applied magnetic fields. Knight shift is defined as $K_{\rm exp}$ = ($B_{\rm int}$ $-$ $B_{\rm ext}$)/$B_{\rm ext}$, where $B_{\rm int}$ and $B_{\rm ext}$ are the internal and externally applied magnetic fields, respectively. $K_{\rm exp}$ shows a sharp changes across $T_{{\rm 1}}^{*}$ and $T_{{\rm 2}}^{*}$, which indicates the change of local magnetic susceptibility with two characteristic temperatures.\\ 

\textbf{Analysis of ZF-${\mu}$SR data under pressure}: As an example, in the Extended data Figure 5 is displaying the zero-field 
${\mu}$SR spectra, recorded at $p$ = 1.07 GPa for various temperatures. The experimental data were analyzed by separating the 
$\mu$SR signal on the sample (s) and the pressure cell (pc) contributions \cite{Khasanov_SciRep_2015,GuguchiaNbSe2}:
 \begin{equation}
A_0  P(t) = A_{\rm s} P_{\rm s}(t) + A_{\rm pc} P_{\rm pc}(t).
 \label{eq:P(t)}
\end{equation}
Here $A_0$ is the initial asymmetry of the muon-spin ensemble, and  $A_{\rm s}$ ($A_{\rm pc}$) and $P_{\rm s}(t)$
[$P_{\rm pc}(t)$] are the asymmetry  and the time evolution of the muon-spin polarization for muons stopped inside the
sample (pressure cell), respectively. The response of the pressure cell [$P_{\rm pc}(t)$] was studied in separate set
of experiments.

The sample contribution includes both, the nuclear moment and an additional exponential relaxation $\Gamma$ caused by
appearance of spontaneous magnetic fields:
\begin{equation}
P_{\rm s}^{\rm ZF}(t) = P_{ZF}^{GKT}(t) e^{-\Gamma t}.
 \label{eq:ZF-P_sample}
\end{equation}
Here $P_{ZF}^{GKT}(t)$ is the Gaussian Kubo-Toyabe (GKT) relaxation function (see Eq. 1) describing
the magnetic field distribution created by the nuclear magnetic moments \cite{Toyabe}.
Fits of Eq.~\ref{eq:P(t)} to the ZF-$\mu$SR pressure data were performed globally. The ZF-$\mu$SR time-spectra taken at
each particular pressure ($p=0.16$, 0.59, 1.07, 1.53, and 1.89~GPa) were fitted simultaneously with  $A_{\rm s}$,
$A_{\rm pc}$, and $\sigma_{\rm GKT}$ as common parameters, and $\lambda$ an individual parameter for each particular
data set. The fits were limited to $T\simeq 150$~K, {\it i.e.} up to the temperature where the nuclear contribution of
RbV$_3$Sb$_5$ remains constant ($\sigma_{\rm GKT}\simeq const$, see Fig.~1e).\\

\textbf{Time-reversal symmetry-breaking charge orders under pressure}: Here we show (see Extended Data Figure 6a-f) the evolution of the two time-reversal symmetry-breaking transition temperatures $T_{{\rm 1}}^{*}$ and $T_{{\rm 2}}^{*}$ with the application of hydrostatic pressure. Two step time-reversal symmetry-breaking transition is clearly observed under the pressures of $p$ = 0.16 GPa and 0.59 GPa. At 1 GPa, these two transitions become indistinguishable and above 1 GPa we see only transition at $T_{{\rm 2}}^{*}$, which decreases upon further increasing the pressure. Extended Data Figure 6f shows the pressure evolution of $T_{{\rm 1}}^{*}$ and $T_{{\rm 2}}^{*}$, extracted from ${\mu}$SR results, and of previously reported charge order temperature $T_{{\rm co,1}}$ \cite{NNWang}. The value of $T_{{\rm co,2}}$ \cite{MWenzel} at ambient pressure is also shown. This phase diagram suggests that two time-reversal symmetry-breaking state turn into single time-reversal symmetry-breaking state at ${\sim}$ 1.5 GPa, above which $T_{{\rm 2}}^{*}$ shows faster suppression and follows the phase line of the charge order. This phase diagram confirms the charge orders being origin for the time-reversal symmetry-breaking in RbV$_{3}$Sb$_{5}$.\\ 

\textbf{Macroscopic superconducting properties under pressure}: The temperature dependence of the AC-susceptibility $\chi_{\rm AC}$ for various pressures, shown in Extended Data Figures 7a and b for the polycrystalline samples of RbV$_{3}$Sb$_{5}$ and KV$_{3}$Sb$_{5}$, indicates a strong diamagnetic response and sharp superconducting transitions in both samples. This points to the high quality of the samples and providing evidence for bulk superconductivity in these polycrystalline samples.\\

\textbf{Analysis of ${\lambda}(T)$}: ${\lambda}$($T$) was calculated within the local (London) approximation (${\lambda}$ ${\gg}$ ${\xi}$) by the following expression \cite{Bastian,Tinkham,Brandt}: 
\begin{equation}
\frac{\lambda^{-2}(T,\Delta_{0,i})}{\lambda^{-2}(0,\Delta_{0,i})}= 1+\frac{1}{\pi}\int_{0}^{2\pi}\int_{\Delta_i(_{T,\varphi})}^{\infty}(\frac{\partial f}{\partial E})\frac{EdEd\varphi}{\sqrt{E^2-\Delta_i(T,\varphi)^2}},
\end{equation}
where $f=[1+\exp(E/k_{\rm B}T)]^{-1}$ is the Fermi function, ${\varphi}$ is the angle along the Fermi surface, and ${\Delta}_{i}(T,{\varphi})={\Delta}_{0,i}{\Gamma}(T/T_{\rm c})g({\varphi}$)
(${\Delta}_{0,i}$ is the maximum gap value at $T=0$). 
The temperature dependence of the gap is approximated by the expression 
${\Gamma}(T/T_{\rm c})=\tanh{\{}1.82[1.018(T_{\rm c}/T-1)]^{0.51}{\}}$,\cite{carrington} 
while $g({\varphi}$) describes the angular dependence of the gap and it is replaced by 1 for both an $s$-wave and an $s$+$s$-wave gap,
${\mid}\cos(2{\varphi}){\mid}$ for a $d$-wave gap, and ${\mid}\cos(6{\varphi}){\mid}$ for a $f$-wave gap.\\

For RbV$_{3}$Sb$_{5}$ and KV$_{3}$Sb$_{5}$, the $\lambda^{-2}(T)$ data above $p_{\rm max-Tc}$ are analysed using two $s$-wave gaps. 
At pressure below $p_{\rm max-Tc}$, the combination of dominant nodal ${\mid}\cos(6{\varphi}){\mid}$-gap and one $s$-wave gap is used.\\

\textbf{Analysis of the temperature dependence of the penetration depth for the single crystals RbV$_{3}$Sb$_{5}$ and KV$_{3}$Sb$_{5}$ at ambient pressure}: ${\lambda}_{eff}^{-2}(T)$ at ambient pressure were analyzed within the framework of quasi-classical Eilenberger weak-coupling formalism, where the temperature dependence of the gaps was obtained by solving self-consistent coupled gap equations. This method is described in details elsewhere \cite{Prozorov_SST_2006,Khasanov_PRL_2007,Khasanov_PRB_2014,Kogan_PRB_1981}, including in our recent paper on KV$_{3}$Sb$_{5}$ \cite{GuguchiaMielke}. The temperature dependence of $\lambda_{ab}^{-2}$ down to 18 mK  in the applied field of 5~mT is shown in Extended Data Figure 8 for RbV$_{3}$Sb$_{5}$ along with the KV$_{3}$Sb$_{5}$ data. A well pronounced two step behaviour is observed in RbV$_{3}$Sb$_{5}$, similar to KV$_{3}$Sb$_{5}$ \cite{GuguchiaMielke} which was explained with two gap superconductivity with very weak interband coupling (0.001-0.005) and strong electron-phonon coupling. The interband coupling is extremely small which is sufficient to have same values of $T_{{\rm c}}$ for different bands but still shows the two step temperature behaviour of the penetration depth \cite{Kogan}. The $\lambda_{ab}^{-2}(T)$ for both (Rb,K)V$_{3}$Sb$_{5}$ are well described by one constant gap and one dominant angle-dependent ${\mid}\cos(6{\varphi}){\mid}$-gap, indicating the presence of gap nodes. Upon increasing pressure two step behaviour gets smoothed out, but angle-dependent gap becomes more dominant and persists all the way up to $p_{\rm max-Tc}$ $\simeq$ $p_{\rm cr,co}$ $\simeq$ 1.5 GPa and 0.5 GPa for RbV$_{3}$Sb$_{5}$ and KV$_{3}$Sb$_{5}$, respectively. Above the pressure $p_{\rm max-Tc}$. At pressures above $p_{\rm max-Tc}$, the $\lambda^{-2}(T)$ is described by constant gaps.\\



\textbf{TF-${\mu}$SR spectra for RbV$_{3}$Sb$_{5}$ and KV$_{3}$Sb$_{5}$}: Extended Data Figures.~9$\bf{a}$ and $\bf{b}$ show the TF-${\mu}$SR spectra, measured near ambient pressure above and below the superconducting transition temperature $T_{{\rm c}}$ for RbV$_{3}$Sb$_{5}$ and KV$_{3}$Sb$_{5}$, respectively. Extended Data Figures.~9$\bf{c}$ and $\bf{d}$ show the TF-${\mu}$SR spectra, measured above and below the superconducting transition temperature $T_{{\rm c}}$ for RbV$_{3}$Sb$_{5}$ at $p$ = 1.85 GPa and for KV$_{3}$Sb$_{5}$ at $p$ = 1.1 GPa, respectively. In order to obtain well ordered vortex lattice, the measurements were done after field cooling the sample from above $T_{\rm c}$. Above $T_{{\rm c}}$, the oscillations show a damping essentially due to the random local fields from the nuclear magnetic moments. Below $T_{{\rm c}}$ the damping rate increases with decreasing temperature due to the presence of a nonuniform local magnetic field distribution as a result of the formation of a flux-line lattice in the superconducting state. Figures~9$\bf{c}$ and $\bf{d}$ show that damping in the superconducting state significantly increases upon application of hydrostatic pressure.\\

\textbf{Analysis of TF-${\mu}$SR data under pressure}: The TF ${\mu}$SR data were analyzed by using the following functional form:\cite{Bastian}
\begin{equation}
\begin{aligned}
P(t)=A_s\exp\Big[-\frac{(\sigma_{\rm sc}^2+\sigma_{\rm nm}^2)t^2}{2}\Big]\cos(\gamma_{\mu}B_{\rm int,s}t+\varphi) \\
 + A_{\rm pc}\exp\Big[-\frac{\sigma_{\rm pc}^2t^2}{2}\Big]\cos(\gamma_{\mu}B_{\rm int,pc}t+\varphi), 
\end{aligned}
\end{equation}
 Here $A_{\rm s}$ and $A_{\rm pc}$  denote the initial assymmetries of the sample and the pressure cell, respectively. ${\varphi}$ is the initial phase of the muon-spin ensemble and $B_{\rm int}$ represents the internal magnetic field at the muon site. The relaxation rates ${\sigma}_{\rm sc}$ and ${\sigma}_{\rm nm}$ characterize the damping due to the formation of the FLL in the superconducting state and of the nuclear magnetic dipolar contribution, respectively. In the analysis ${\sigma}_{\rm nm}$ was assumed to be constant over the entire temperature range and was fixed to the value obtained above $T_{\rm c}$ where only nuclear magnetic moments contribute to the muon depolarization rate ${\sigma}$. The Gaussian relaxation rate, ${\sigma}_{\rm pc}$, reflects the depolarization due to the nuclear moments of the pressure cell. The width of the pressure cell signal increases below $T_{c}$. This is due to the influence of the diamagnetic moment of the superconducting sample on the pressure cell, leading to the temperature dependent ${\sigma}_{\rm pc}$ below $T_{c}$. In order to consider this influence we assume the linear coupling between ${\sigma}_{\rm pc}$ and the field shift of the internal magnetic field in the superconducting state:\\
 
${\sigma}_{\rm pc}$($T$) = ${\sigma}_{\rm pc}$($T$ ${\textgreater}$ $T_{\rm c}$) + $C(T)$(${\mu}_{\rm 0}$$H_{\rm int,NS}$ - ${\mu}_{\rm 0}$$H_{\rm int,SC}$), \\ 

where  ${\sigma}_{\rm pc}$($T$ ${\textgreater}$ $T_{\rm c}$) = 0.25 ${\mu}$$s^{-1}$ is the temperature independent Gaussian relaxation rate. ${\mu}_{\rm 0}$$H_{\rm int,NS}$ and ${\mu}_{\rm 0}$$H_{\rm int,SC}$ are the internal magnetic fields measured in the normal and in the superconducting state, respectively. As indicated by the solid lines in Extended data Figs.~9a-d the ${\mu}$SR data are well described by Eq.~(5).\\

\textbf{Theoretical model for the nodal-to-nodeless transition}: 
Here we present more details about our theoretical model for the nodal-to-nodeless transition assuming that the ``pure" superconducting state is the time-reversal symmetry-breaking chiral $d_{x^2-y^2} + id_{xy}$ state. The kagome lattice has point group D$_{6h}$ and the $d$-wave order parameter has two degenerate components corresponding to the $\Delta_{d_{x^2-y^2}}=\Delta_1$ and the $\Delta_{d_{xy}}=\Delta_{2}$ order parameters. The combined order parameter,
\begin{equation}
	\boldsymbol{\Delta} = \begin{pmatrix}
		\Delta_{1} \\
		\Delta_{2}
	\end{pmatrix}\,,
\end{equation}
transforms as the $E_{2g}$ irreducible representation (irrep) of the point group. Writing  $\Delta_1 = \Delta_0 \cos \theta e^{i\phi_1}$ and $\Delta_2 = \Delta_0 \sin \theta e^{i\phi_2}$ the Landau free-energy expansion to quartic order is
\begin{equation}
	\mathcal{F}_{\rm SC} = \alpha \Delta_0^2 + \beta_1 \Delta_0^4 - \beta_2 \Delta_0^4  \sin^2  2\theta \sin^2 \phi \,,  \label{aux1}
\end{equation}
where $\phi = \phi_1 - \phi_2$ is the relative phase between the two order parameters. From here we see immediately the well-known result that for $\beta_2 >0$ a relative phase of $\phi= \pm \pi/2$ is preferred, whereas for $\beta_2 <0$, a relative phase of $\phi=0,\pi$ is selected~\cite{Sigrist1991}. Hereafter we will focus on the $\beta_2 >0$ case and take $\beta_1 > \beta_2$ in order for the free energy to be bounded. In this situation, the $\phi=\pm \pi/2$ phase is selected, implying time-reversal symmetry breaking and a nodeless pairing state.\\

We now consider what happens inside the charge-ordered (CO) state. Some of the proposed $2 \times 2 \times 2$ charge-order configurations, such as the tri-hexagonal, Star of David, and superimposed tri-hexagonal Star of David phases \cite{MHChristensen}, are triple-$\mathbf{Q}_M$/triple-$\mathbf{Q}_L$ states that preserve the D$_{6h}$ point group of the kagome lattice. Here, $\mathbf{Q}_M$ and $\mathbf{Q}_L$ refer to the wave-vectors $(\frac{1}{2}, \,\frac{1}{2}, \, 0)$  and $(\frac{1}{2}, \,\frac{1}{2}, \, \frac{1}{2})$ of the Brillouin zone. In these cases, because $\boldsymbol{\Delta}$ continues to transform as the two-dimensional $E_{2g}$ irrep, the superconducting state is expected to remain chiral and nodeless. \\

\begin{table*}
	\begin{tabular}{c|c|c|c|c}
		Solution & $\phi_{\ast}$ & $\theta_{\ast}$ & $\Delta_{0,\ast}$ & $\mathcal{F}_{\ast}$ \\
		\hline
		Disordered & $-$ & $-$ & $0$ & $0$  \\
		Nodal (1) & - & $\frac{\pi}{2}$ & $\sqrt{\frac{-\alpha - \lambda \Delta_{\rm CO}^2}{2\beta_1}}$ & $-\frac{(\alpha - \lambda\Delta_{\rm CO}^2)^2}{4\beta_1}$ \\
		Nodeless (2) & $\frac{\pi}{2}$ & $\frac{1}{2}\arctan \left[\frac{\sqrt{4\beta_2^2 \Delta_0^2 - \lambda^2 \Delta_{\rm CO}^4}}{-\lambda\Delta_{\rm CO}^2} \right]$ & $\sqrt{\frac{-\alpha}{2(\beta_1 - \beta_2)}}$ & $-\frac{\alpha^2}{4(\beta_1 - \beta_2)} - \frac{\lambda^2 \Delta_{\rm CO}^4}{4\beta_2}$
	\end{tabular}
	\caption{\label{tab:summary}Comparison between different extrema of the free energy in Eq.~\eqref{eq:free_energy_full}. An additional constraint on the parameters arise as a consequence of the square-root function appearing in the solution for $\theta$.}
\end{table*}

However, other proposed $2 \times 2 \times 2$ CDW phases break the threefold rotational symmetry of the lattice, implying that $\Delta_{1}$ and $\Delta_{2}$ no longer onset at the same temperature. This is the case of the so-called staggered tri-hexagonal and staggered Star of David phases  \cite{MHChristensen}, which are double-$\mathbf{Q}_L$/single-$\mathbf{Q}_M$ states. In this case, a composite quantity transforming as the $E_{2g}$ irrep of the point group can be constructed from the order parameters of the CO state:
\begin{equation}
	 \begin{pmatrix}
		M_1^2 + M_3^2 - 2 M_2^2  \\
		\sqrt{3} (M_3^2 - M_1^2)
	\end{pmatrix}\,, \label{aux2}
\end{equation}
where $M_i$, with $i=1,2,3$, denote the CO order parameter associated with each wave-vector in the star of $\mathbf{Q}_M$. While $\boldsymbol{\Delta}$ above also transforms as $E_{2g}$, it cannot be combined with the above composite in the free energy, as it is not gauge-invariant. Nevertheless, we can construct a composite superconducting order parameter combination that is gauge invariant and still transforms as the $E_{2g}$ irrep:

\begin{equation}
	\begin{pmatrix}
		|\Delta_1|^2 - |\Delta_2|^2 \\
		-\Delta_1 \Delta_2^{\ast} - \Delta_1^{\ast}\Delta_2
	\end{pmatrix}\,, \label{aux3}
\end{equation}

The ``scalar product" between the composites (\ref{aux2}) and (\ref{aux3}) is now gauge-invariant and transforms trivially under the point group operations and, thus, an allowed term in the free energy expansion. Since we are interested in the fate of the superconducting state inside the CO state, we consider for concreteness and without loss of generality, the particular configuration $M_1 = M_3 = \Delta_{\rm CO}/\sqrt{2}$, $M_2 = 0$. The full expression for the free energy then reads

\begin{align}
	\mathcal{F} & = \alpha \Delta_0^2 + \beta_1 \Delta_0^4 - \beta_2 \Delta_0^4  \sin^2  2\theta \sin^2 \phi  \nonumber \\
	& + \lambda \Delta_0^2  \Delta_{\rm CO}^2 \cos 2 \theta \,, \label{eq:free_energy_full}
\end{align}

where $\lambda$ is a coupling constant, assumed hereafter to be positive. Minimization of the free energy yield two possible minima, whose free energies are given by:
\begin{align}
	\mathcal{F}_{1} &= - \frac{(\alpha - \lambda \eta_1)^2}{4\beta_1} \\
	\mathcal{F}_{2} &= -\frac{\alpha^2}{4(\beta_1-\beta_2)} - \frac{\lambda^2 \eta_1^2}{4\beta_2}\,.
\end{align}
As summarized in Table~\ref{tab:summary}, solution 1 corresponds to a superconducting state where only $\Delta_1$ is non-zero, resulting in a nodal state, since the gap $\Delta_1$ must vanish along the $k_x = \pm k_y$ directions. In contrast, in solution $2$, both $\Delta_1$ and $\Delta_2$ are non-zero. Although they have different magnitudes and their relative phase is no longer $\pm \pi/2$, the total gap function is always finite, implying that solution 2 it is a nodeless state. \\

The solution that minimizes the free energy depends on the values of $\alpha$ and $\Delta_{\rm CO}^2$. The nodeless state (solution 2) takes place as long as the following condition is met:
\begin{equation}
	\Delta_{\rm CO}^2 \leq  -\frac{\alpha \beta_2}{\lambda(\beta_1-\beta_2)} \,,
\end{equation}
which arises from enforcing the argument of square root in the expression for the angle $\theta_{\ast}$ to be positive (see Table~\ref{tab:summary}). When the constraint is saturated, i.e. for
\begin{equation}
	\Delta_{\rm CO}^2 = -\frac{\alpha \beta_2}{\lambda(\beta_1-\beta_2)}\,,
\end{equation}
we have $\mathcal{F}_{1} = \mathcal{F}_{2}$. For larger values of $\Delta_{\rm CO}$, the nodal state (solution 1) is favored. Defining
\begin{equation}
	\alpha = \alpha_0 (t-1)\,,
\end{equation}
where $t=\frac{T}{T_{c,0}}$ is the reduced temperature, we can obtain the transition temperatures for both solutions as a function of $\Delta_{\rm CO}^2$. We find
\begin{align}
	t_{c,\rm{nodal}} &= 1 + \frac{\lambda \Delta_{\rm CO}^2}{\alpha_0}\\
	t_{c,\rm{nodeless}} &= 1 - \frac{\lambda \Delta_{\rm CO}^2}{\alpha_0\beta_2}(\beta_1 - \beta_2)\,.
\end{align}

Hence, for a finite CO order parameter, the nodal state onsets first, followed by a transition at lower temperatures to a nodeless state. Accordingly, for a fixed temperature, there is a nodeless to nodal transition upon increasing $\Delta_{\rm CO}$. This is illustrated in Extended Data Fig. \ref{fig_th}, which shows $t_{c,\rm{nodal}}$ and $t_{c,\rm{nodeless}}$ as a function of $\Delta_{\rm CO}$. The specific parameters used in making the figure were: $\alpha_0=0.1$, $\beta_1=1$, $\beta_2=0.4$, and $\lambda=0.25$. \\

It is important to emphasize that, even in the nodeless state, the minimum gap value can be very small. To show that, we consider the full gap function
\begin{equation}
	\Delta_{\rm tot} = f_{k_x^2 - k_y^2} \Delta_0 \cos \theta + f_{k_x k_y} \Delta_0 \sin \theta e^{i\phi}\,, 
\end{equation}
where $f_{k_x^2-k_y^2}$ and $f_{k_x k_y}$ are form factors that vanish at $k_x = \pm k_y$ and $k_x = 0$ or $k_y=0$, respectively. Here, $\theta$ and $\phi$ are functions of $\Delta_{\rm CO}$ and are given in Table~\ref{tab:summary}.  Evaluating
\begin{equation}
	\min \left| \frac{\Delta_{\rm tot}}{\Delta_0} \right|\,,
\end{equation}
as a function of $\Delta_{\rm CO}$ gives the inset of Extended Data Fig.\ref{fig_th}, which was obtained using the same parameters as above and setting the reduced temperature to $t=0.4$. As expected, the gap minimum vanishes continuously across the nodeless to nodal transition. \\

\section{Acknowledgments}~
The ${\mu}$SR experiments were carried out at the Swiss Muon Source (S${\mu}$S) Paul Scherrer Institute, Villigen, Switzerland. 
Z.G. acknowledges useful discussions with Dr. Robert Johann Scheuermann. M.H.C. was supported by the Carlsberg foundation. M.Z.H. acknowledges visiting scientist support from IQIM at the California Institute of Technology. Experimental work at Princeton University was supported by the Gordon and Betty Moore Foundation (GBMF4547 and GBMF9461; M.Z.H.) and the material characterization is supported by the US Department of Energy under the Basic Energy Sciences programme (grant no. DOE/BES DE-FG-02-05ER46200). Y.S. acknowledges the National Natural Science Foundation of China ( U2032204) and the Strategic Priority Research Program (B) of the Chinese Academy of Sciences (No. XDB33000000). H.C.L. was supported by Ministry of Science and Technology of China (Grant No. 2018YFE0202600), Beijing Natural Science Foundation (Grant No. Z200005). The work of R.G. was supported by the Swiss National Science Foundation (SNF-Grant No.\ 200021-175935). R.M.F (phenomenological modeling) was supported by the Air Force Office of Scientific Research under award number FA9550-21-1-0423. 


\pagebreak

\renewcommand{\figurename}{Extended Data Figure}

\renewcommand\thefigure{\arabic{figure}}
\setcounter{figure}{0}

\begin{figure*}[t!]
\includegraphics[width=1.0\linewidth]{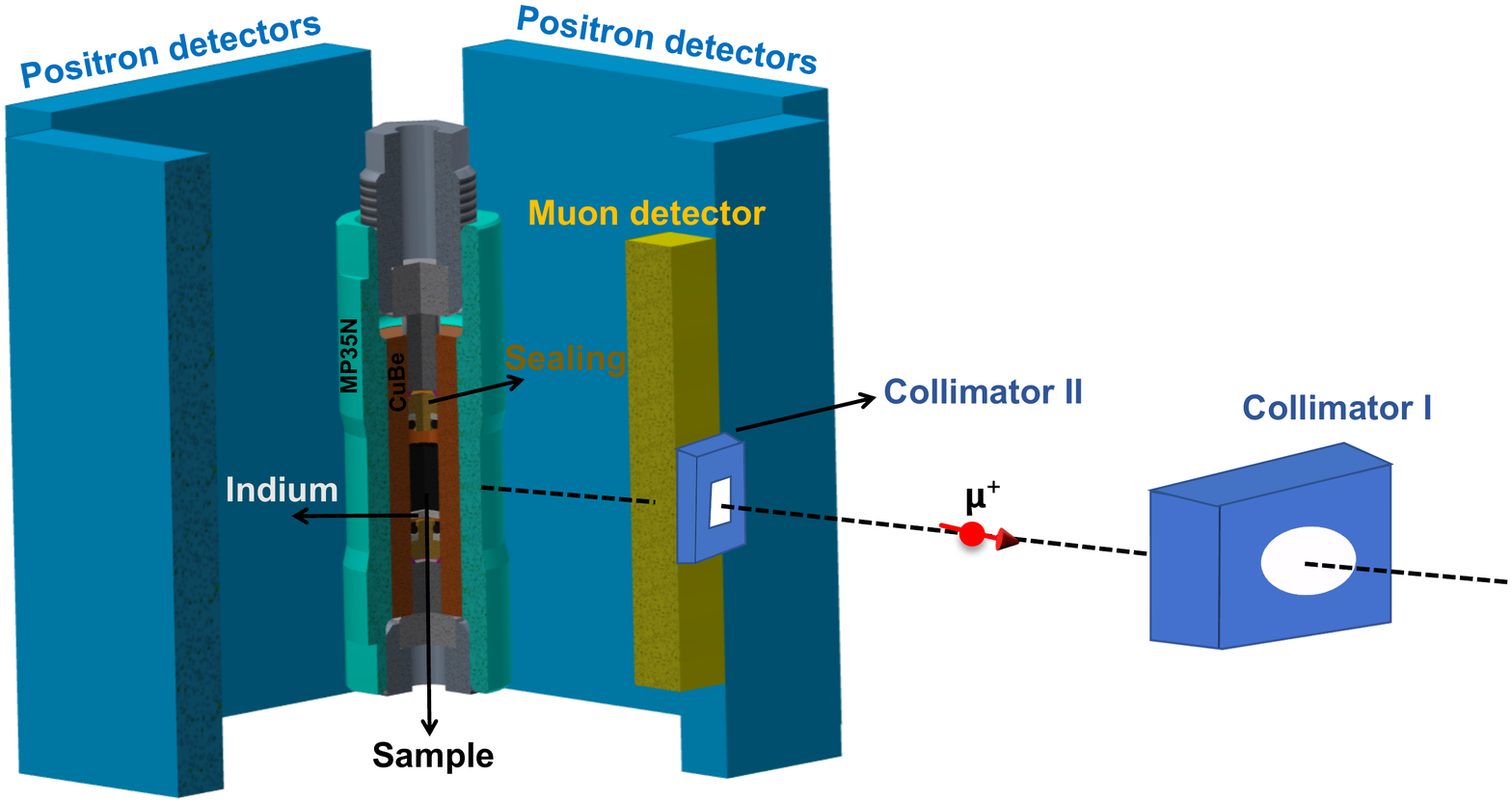}
\vspace{-2cm}
\caption{ \textbf{Pressure cell for ${\mu}$SR.}
Fully assembled typical double-wall piston-cylinder type of pressure cell used in our ${\mu}$SR experiments.  The schematic view of the positron and muon detectors at the GPD spectrometer are also shown. In reality, each positron detector consists of three segments. The collimators reduce the size of the incoming muon beam.}
\label{fig1}
\end{figure*}

\begin{figure*}[t!]
\includegraphics[width=1.0\linewidth]{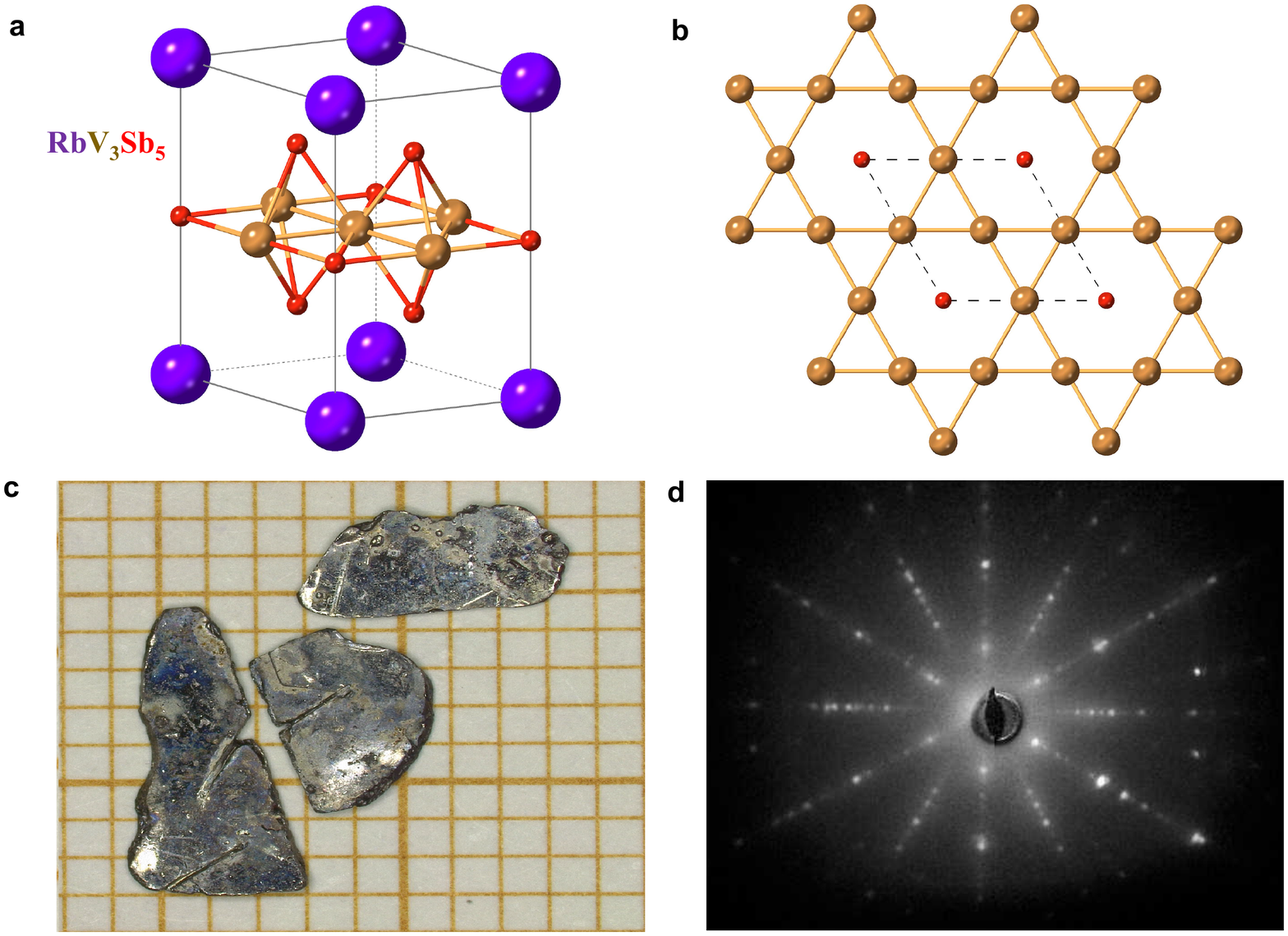}
\vspace{-0.7cm}
\caption{ \textbf{Crystal structure of RbV$_{3}$Sb$_{5}$.}
Three dimensional representation (a) and top view (b) of the atomic structure of RbV$_{3}$Sb$_{5}$.
(c) An optical microscope images of several single crystals of RbV$_{3}$Sb$_{5}$ on millimeter paper. The hexagonal symmetry is immediately apparent. (d) Laue X-ray diffraction image of the single crystal sample of RbV$_{3}$Sb$_{5}$, oriented with the $c$-axis along the beam.}
\label{fig1}
\end{figure*}

\begin{figure*}[t!]
\includegraphics[width=0.9\linewidth]{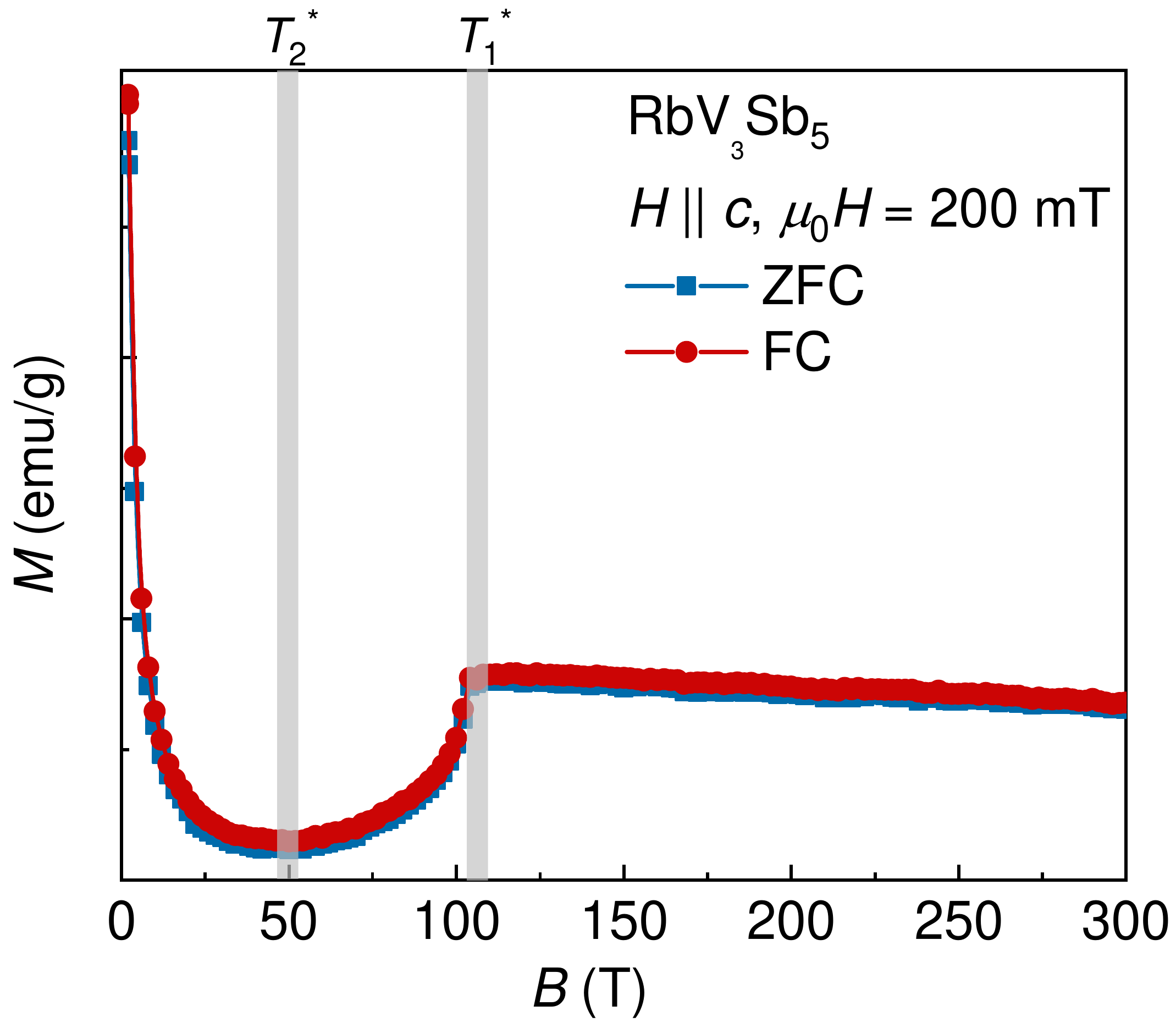}
\vspace{0cm}
\caption{ \textbf{Bulk magnetization for RbV$_{3}$Sb$_{5}$.}
The temperature dependence of magnetic susceptibility of RbV$_{3}$Sb$_{5}$ above 1.8 K. The vertical grey lines mark the concomitant 
time-reversal symmetry-breaking and charge ordering temperatures $T_{{\rm 1}}^{*}$ = $T_{\rm CDW,1}$ ${\simeq}$ 110 K, 
$T_{{\rm 2}}^{*}$ = $T_{\rm CDW,2}$ ${\simeq}$ 50 K.} 
\label{fig1}
\end{figure*}


\begin{figure*}[t!]
\includegraphics[width=0.9\linewidth]{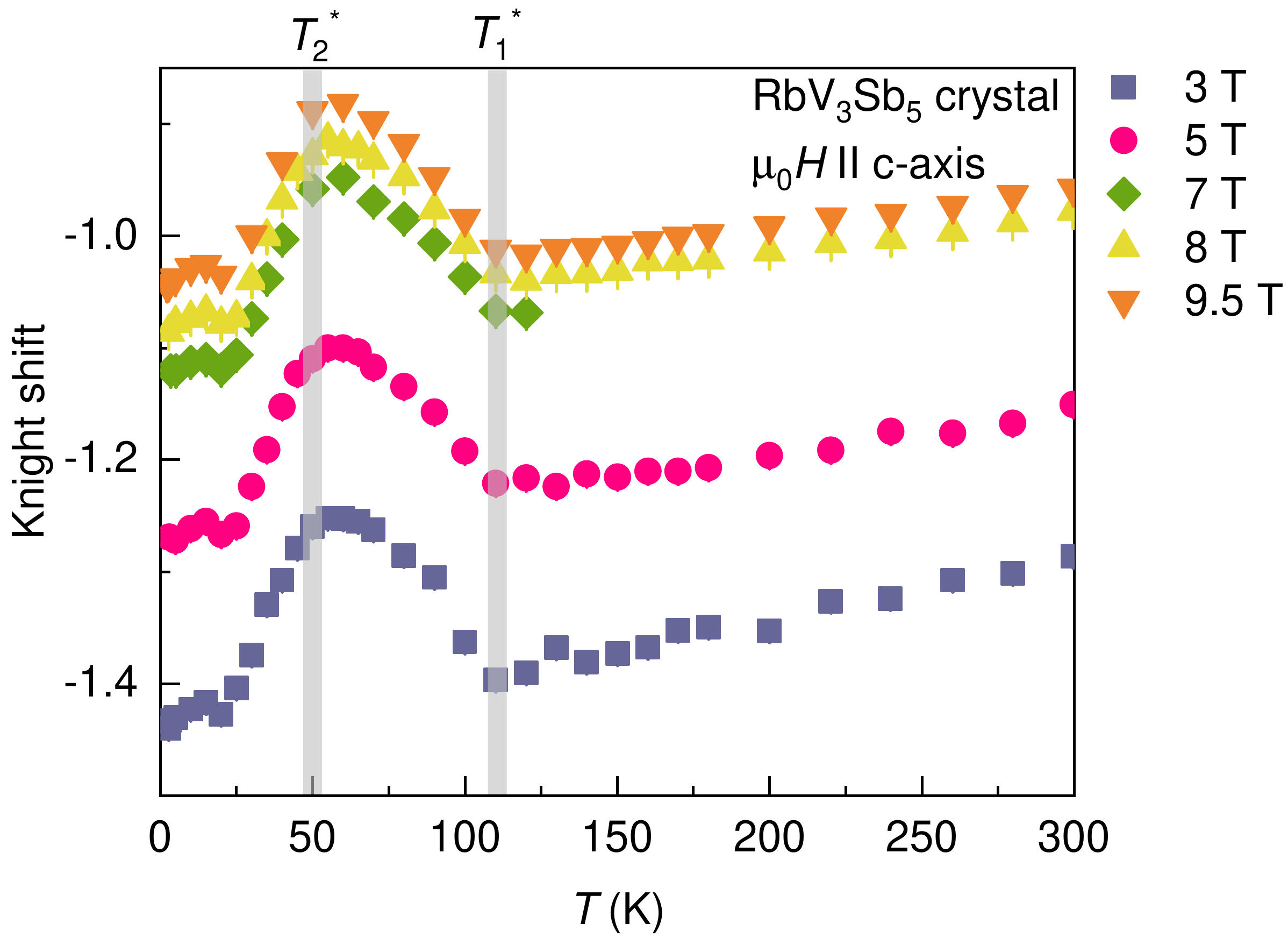}
\vspace{0cm}
\caption{ \textbf{Knight shift for RbV$_{3}$Sb$_{5}$.}
The temperature dependence of the Knight shift for the single crystal of RbV$_{3}$Sb$_{5}$, measured at various magnetic fields applied along the $c$-axis.}
\label{fig1}
\end{figure*}

\begin{figure*}[t!]
\includegraphics[width=0.7\linewidth]{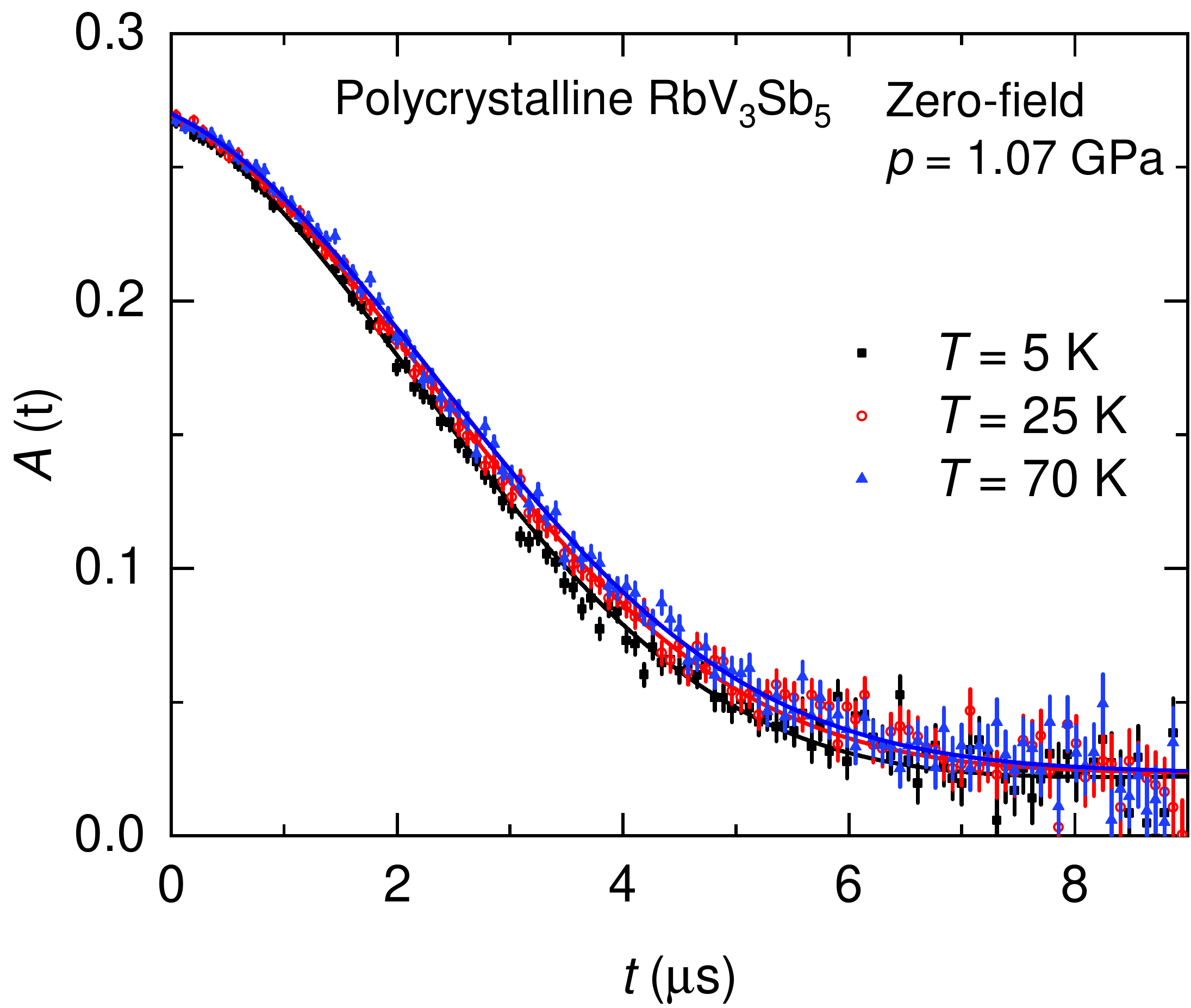}
\vspace{0cm}
\caption{ \textbf{Zero-field spectra of RbV$_{3}$Sb$_{5}$ under pressure.}
The ZF-${\mu}$SR time spectra for the polycrystalline sample of RbV$_{3}$Sb$_{5}$, recorded at various temperatures under the applied pressure of $p$ = 1.07 GPa. The solid lines in panel a represent fits to the data by means of Eq. 3.}
\label{fig1}
\end{figure*}

\begin{figure*}[t!]
\centering
\includegraphics[width=1.0\linewidth]{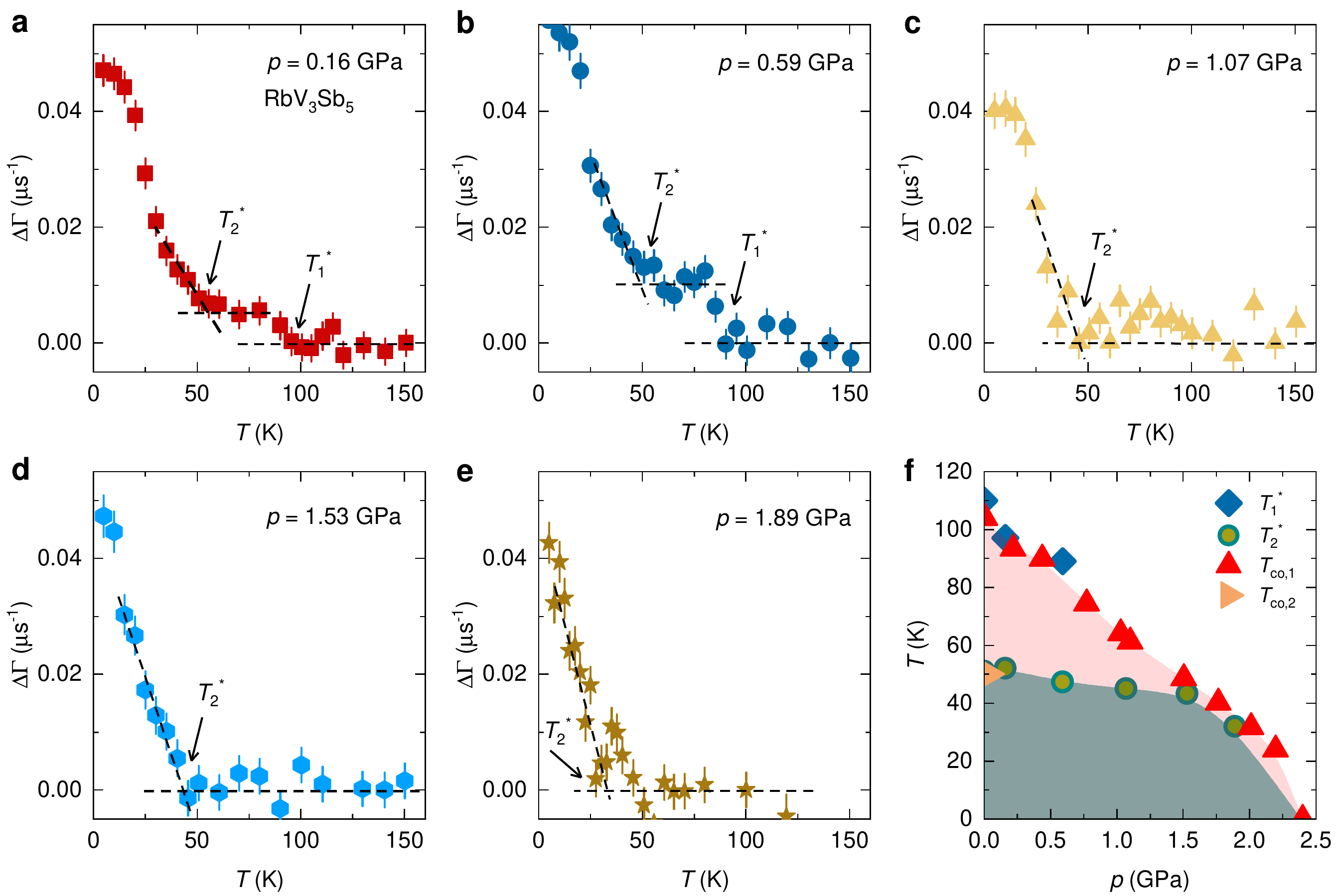}
\vspace{-0cm}
\caption{ (Color online) \textbf{Pressure evolution of time-reversal symmetry-breaking charge orders in RbV$_{3}$Sb$_{5}$.} 
(a-e) The temperature dependence of the absolute change of the electronic relaxation rate ${\Delta}$${\Gamma}$ = ${\Gamma}$($T$) - ${\Gamma}$($T$  ${\textgreater}$ 150 K)  for the polycrystalline sample of RbV$_{3}$Sb$_{5}$, measured at various pressures. (f) The charge order temperatures $T_{{\rm co,1}}$, $T_{{\rm co,2}}$ (after References \cite{MWenzel}, \cite{NNWang}, \cite{FDu}) and the onset temperatures of the time-reversal symmetry-breaking $T_{{\rm 1}}^{*}$, $T_{{\rm 2}}^{*}$ as a function of pressure.}
\label{fig2}
\end{figure*}

\begin{figure*}[t!]
\includegraphics[width=1.0\linewidth]{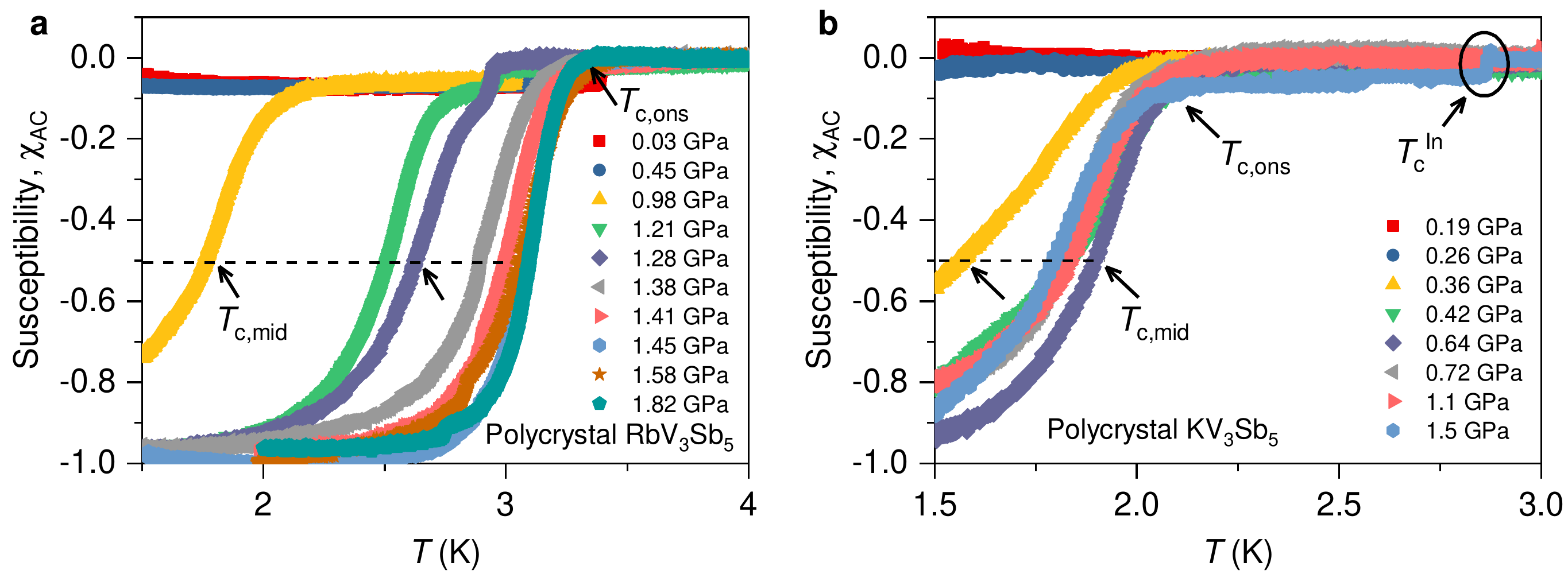}
\vspace{0.5cm}
\caption{ \textbf{Macroscopic superconducting properties under pressure.}
Temperature dependence of the AC susceptibility ${\chi}$ for the polycrystalline samples of RbV$_{3}$Sb$_{5}$ (a) and KV$_{3}$Sb$_{5}$ (b), 
measured at nearly ambient and various applied hydrostatic pressures up to $p$ ${\simeq}$ 1.8 GPa. Arrows mark the onset temperature $T_{\rm c,ons}$ and the temperature $T_{\rm c,mid}$ at which $\chi_{\rm dc}$ = -0.5.}
\label{fig1}
\end{figure*}

\begin{figure*}[t!]
\includegraphics[width=0.7\linewidth]{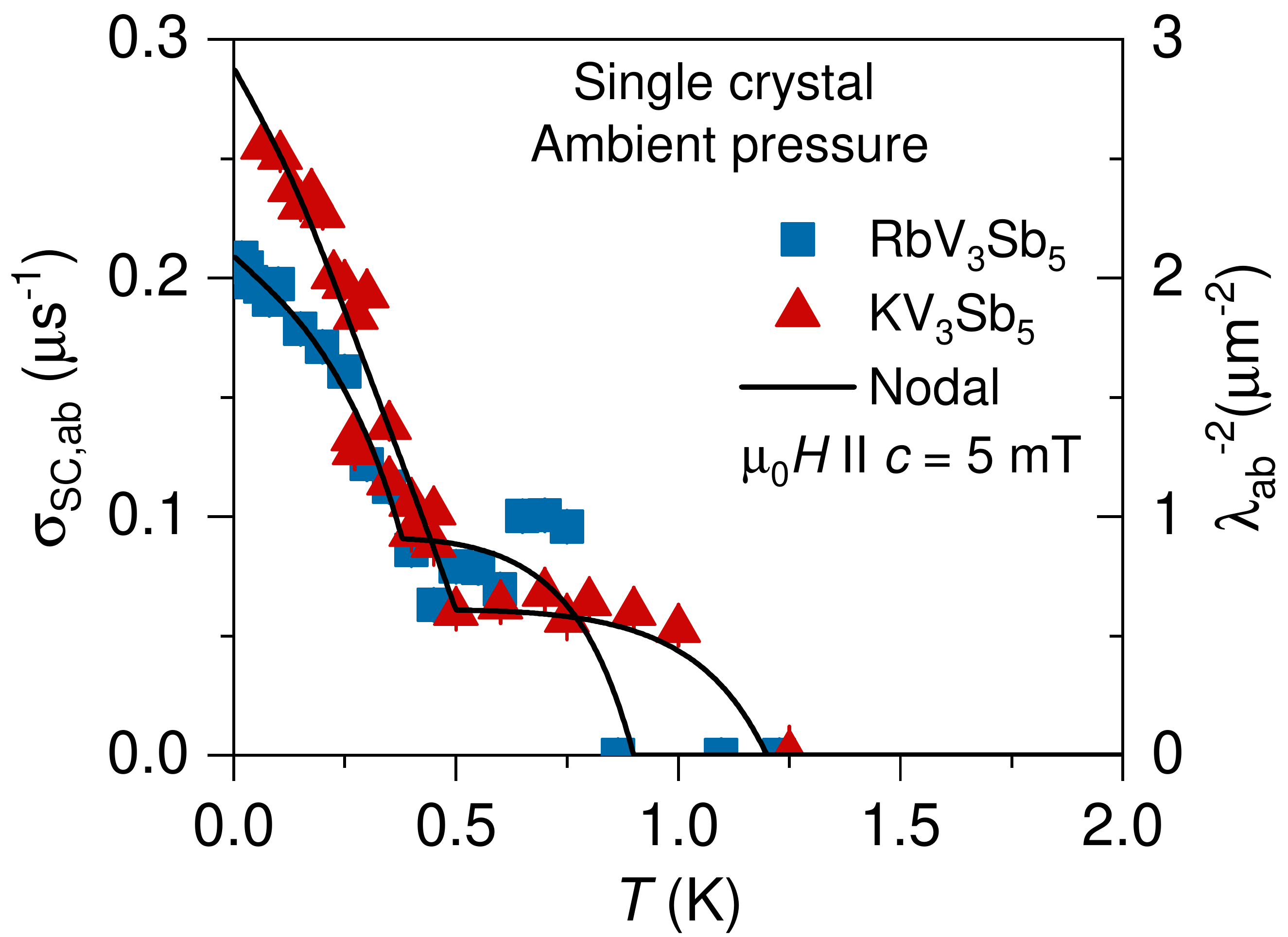}
\vspace{0cm}
\caption{ \textbf{Temperature dependence of the penetration depth at ambient pressure.}
The superconducting muon depolarization rate ${\sigma}_{sc,ab}$ for the single crystals of RbV$_{3}$Sb$_{5}$ and KV$_{3}$Sb$_{5}$  as a function of temperature, measured in 5 mT applied perpendicular to the kagome plane. The solid lines correspond to a model with one constant gap and one dominant angle-dependent gap.}
\label{fig1}
\end{figure*}

\begin{figure*}[t!]
\includegraphics[width=1.0\linewidth]{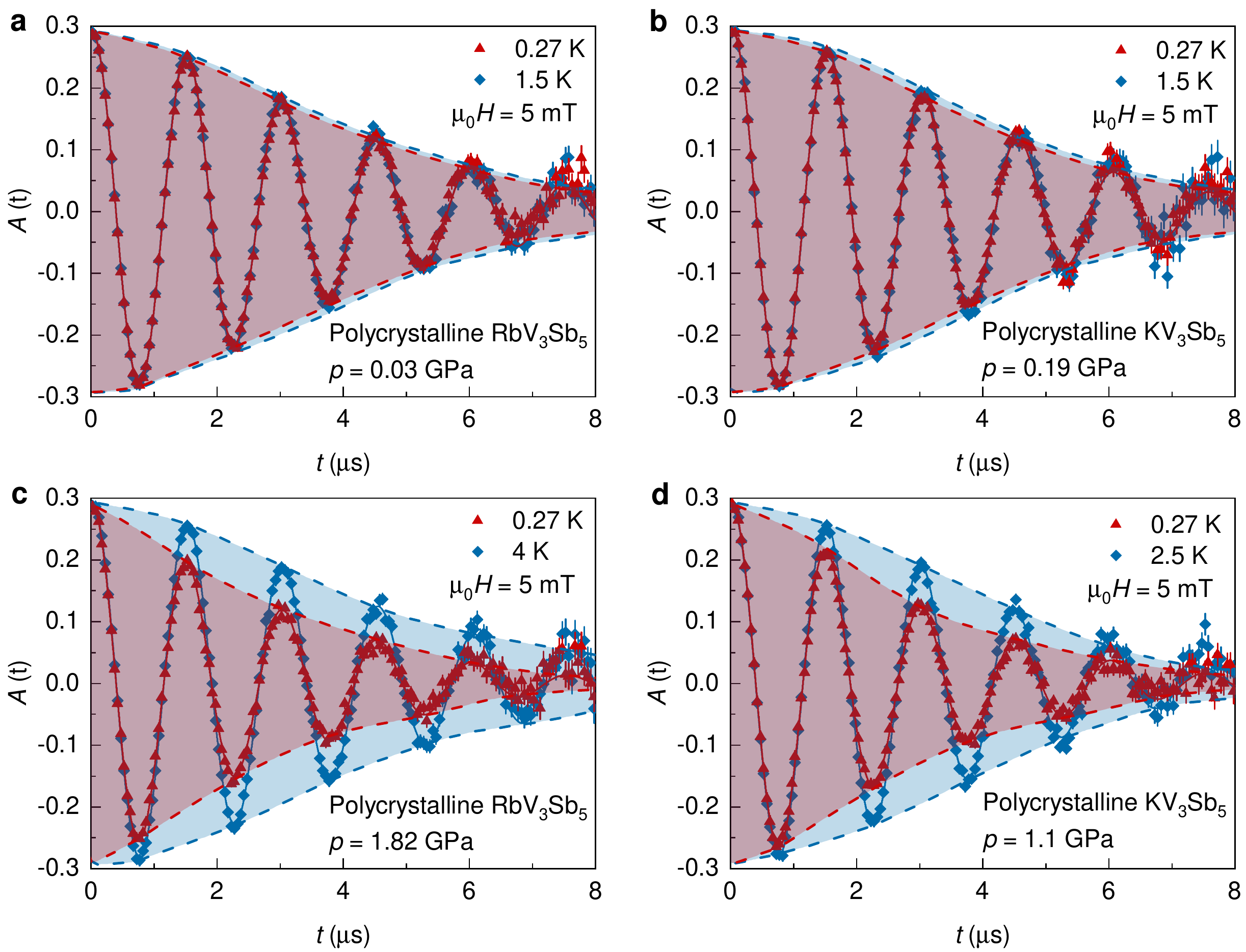}
\vspace{0cm}
\caption{ \textbf{Transverse field ${\mu}$SR spectra.}
The transverse field ${\mu}$SR spectra for RbV$_{3}$Sb$_{5}$ (a,c) and KV$_{3}$Sb$_{5}$ (b,d), obtained 
above and below $T_{\rm c}$ (after field cooling the sample from above $T_{\rm c}$) close to ambient pressure (a and b) and at the maximum applied pressure (c and d). Error bars are the standard error of the mean (s.e.m.) in about 10$^{6}$ events. The error of each bin count $n$ is given by the standard deviation (s.d.) of $n$. The errors of each bin in $A(t)$ are then calculated by s.e. propagation. The solid lines in panel a represent fits to the data by means of Eq. 5. The dashed lines are the guides to the eyes.}
\label{fig1}
\end{figure*}

\begin{figure*}[t!]
\centering
\includegraphics[width=0.8\linewidth]{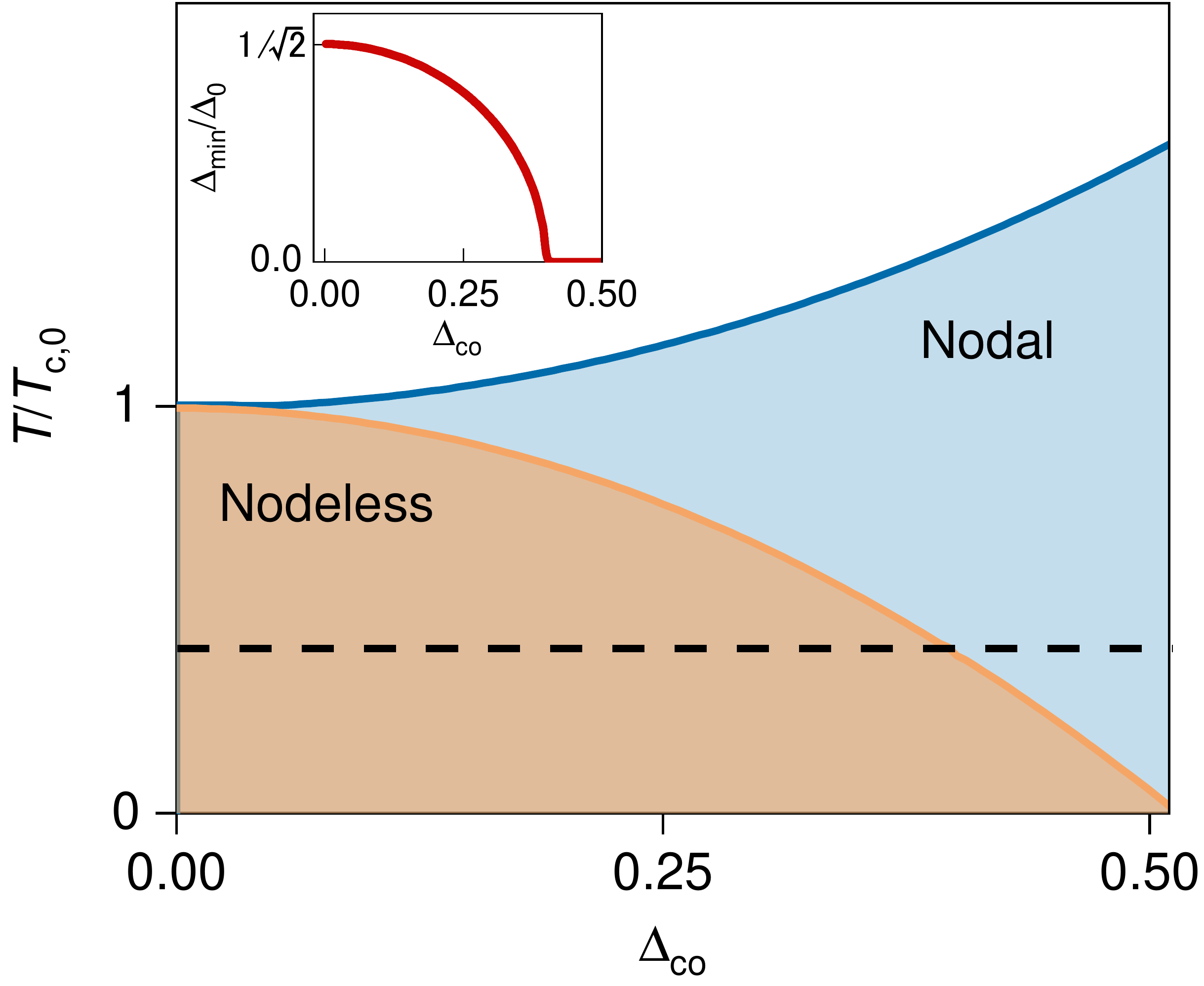}
\vspace{0cm}
\caption{\label{fig:theory_calc} (Color online) \textbf{Calculated superconducting phase diagram.} 
Normalized superconducting critical temperature as a function of the charge order parameter, $\Delta_{\rm co}$. As $\Delta_{\rm co}$ is increased, a transition from nodeless to nodal superconductivity occurs. As evidenced in Fig.~\ref{fig:phase_diagrams}{\bf c} and {\bf d}, the charge order is suppressed by pressure. As pressure is increased, $\Delta_{\rm co}$ is reduced, and the superconducting state goes from nodal to nodeless. The inset shows the minimum gap magnitude as a function of $\Delta_{\rm co}$ plotted along the dashed line in the phase diagram. The transition between nodal and nodeless superconductivity is clearly visible.}
\label{fig_th}
\end{figure*}

\end{document}